\PassOptionsToPackage{table,xcdraw}{xcolor}
\documentclass[sigconf]{acmart}
\AtBeginDocument{%
  \providecommand\BibTeX{{%
    \normalfont B\kern-0.5em{\scshape i\kern-0.25em b}\kern-0.8em\TeX}}}

\usepackage{todonotes}
\usepackage[ruled,vlined,linesnumbered,algo2e]{algorithm2e} 
\usepackage[compatible]{algpseudocode}

\usepackage[framemethod=tikz]{mdframed}
\usepackage{xcolor}
\usepackage{tcolorbox}

\usepackage{caption}
\usepackage{subcaption}

\usepackage{wrapfig,lipsum,booktabs}

\usepackage{graphicx}

\begin{document}


%

\title{Unify and Triumph: Polyglot, Diverse, and Self-Consistent Generation of Unit Tests with LLMs}

\author{Djamel Eddine Khelladi}
\affiliation{%
  \institution{Univ Rennes, Inria, CNRS, IRISA}
  \city{Rennes}
  \country{France}}
\email{djamel-eddine.khelladi@irisa.fr}

\author{Charly Reux}
\affiliation{%
  \institution{Univ Rennes, Inria, CNRS, IRISA}
  \city{Rennes}
  \country{France}}
\email{charly.reux@inria.fr}

\author{Mathieu Acher}
\affiliation{%
  \institution{Univ Rennes, Inria, CNRS, IUF, IRISA}
  \city{Rennes}
  \country{France}}
\email{mathieu.acher@irisa.fr}

\settopmatter{printacmref=false}
\renewcommand\footnotetextcopyrightpermission[1]{}

\acmConference{}{}{}


\newcommand{\polytestname}{\emph{PolyTest}}
\newcommand{\polytest}{\polytestname\xspace}
\newcommand{\red}[1]{{\color{red}{#1}}}

\begin{abstract}

%

Large language model (LLM)-based test generation has gained attention in software engineering, yet most studies evaluate LLMs' ability to generate unit tests in a single attempt for a given language, missing the opportunity to leverage LLM diversity for more robust testing. 
This paper introduces \polytest a novel approach that enhances test generation by exploiting polyglot and temperature-controlled diversity. 
\polytest systematically leverages these properties in two complementary ways: (1) Cross-lingual test generation: Tests are generated in multiple languages at zero temperature and then unified; (2) Diverse test sampling: Multiple test sets are generated within the same language at a higher temperature before unification. 
A key idea is that LLMs can generate diverse yet contradicting tests -- same input, different expected outputs -- across languages and generations. 
\polytest mitigates these inconsistencies by unifying test sets across languages and generations, fostering self-consistency and improving overall test quality.

We evaluate \polytest on \emph{LLama3-70b}, \emph{GPT-4o}, and \emph{GPT-3.5} using the latest version of the dataset EvalPlus that contains curated prompts of coding problems and canonical solutions. 
On 164 problems, we generate tests at temperature 0 for five languages, namely Java, C, Python, JavaScript, and also in a language-agnostic format of input/output in a CSV. 
We also generate tests five times for each of the above languages with a high temperature at 1. 
We perform the union of the tests at each step with our \polytest approach (i.e., between the 5 languages and between the 5 generations per language).  
We observe up to \emph{\textbf{15.42\%}} contradicting tests (in JavaScript with GPT-4o),  \emph{\textbf{7.41\%}} (in C with Llama3-70b), and \emph{\textbf{6.51\%}} (in Java with GPT-3.5).
Results also show that \polytest in both polyglot and temperature-controlled diversity is indeed able to improve the obtained tests w.r.t. all metrics we considered, 
namely number of tests and of passing tests (multiplied up to \emph{\textbf{x2.67}} and \emph{\textbf{x2.85}}), statement and branch coverage (up to \emph{\textbf{+7.9\%}} and \emph{\textbf{+9.01\%}}), and mutation score (up to \emph{\textbf{+11.23\%}}). Overall, \polytest outperforms single-language and single-attempt approaches without requiring on-the-fly execution of every test case, and is particularly beneficial for programming languages where LLMs exhibiting weak performance. 
Finally, \polytest also outperformed Pynguin, as a baseline comparison, in generated/passing tests and mutation score. 
 

\end{abstract}




\maketitle

\section{Introduction}

Large language models (LLMs) have emerged in the field of natural language processing, exhibiting high aptitude to transform and generate textual data. 
 Since their appearance, LLMs have been applied in different domains and tasks in Software Engineering \cite{10109345,10173990,liu2023improving,hou2023large,pearce2022asleep,sobania2022choose,ziegler2022productivity,vaithilingam2022expectation,nguyen2022empirical,doderlein2022piloting,
nathalia2023artificial,yeticstiren2023evaluating,guo2023exploring,fu2023chatgpt,kabir2023empirical,chaaben2023towards,camara2023assessment,AbukhalafHK23}. 

Testing is a crucial part of software development to ensure  quality and correctness of software. However, manually specifying and writing relevant tests is a non-trivial task. 
Hence, an extensive literature emerged on automatic unit test\footnote{For simplicity, we will refer to tests rather than unit tests in the rest of the paper.} generation among which lately LLM-based test generation has attracted attention \cite{wang2024software}. 
In fact, LLMs stand as promising tools for tackling increasingly complex problems and support developers in various tasks of writing, correcting and documenting source code and other artifacts. While there is an extensive empirical assessment of the LLMs capabilities in generating code \cite{10109345,10173990,liu2023improving,hou2023large,pearce2022asleep,sobania2022choose,ziegler2022productivity,vaithilingam2022expectation,nguyen2022empirical,doderlein2022piloting,
nathalia2023artificial,yeticstiren2023evaluating,guo2023exploring,fu2023chatgpt,kabir2023empirical,chaaben2023towards,camara2023assessment,AbukhalafHK23}, there are less works assessing their ability to generate tests \cite{schafer2023empirical,siddiq2024using,baudry2024generative,sapozhnikov2024testspark,li2024large}. 
However, to the best of our knowledge, they only evaluate the capability of LLMs to generate better tests in one shot for a target single language or set of languages, or improve the tests with static analysis, mutation, or repair \cite{gu2024testart,pan2024multi,dakhel2024effective}.

This paper introduces \polytest a novel approach that enhances tests generation by exploiting the diversity of LLMs' output induced by multi-lingual (a.k.a. polyglot) and temperature-control. 
In fact, one of the powerful diversity features of LLMs is their polyglot nature, i.e., trained on multiple languages, and hence, capable of handling tasks across multiple languages. For example, LLMs can 
translate code from a source language to a different target language, and generating code and tests for multiple languages, such as Java, C, Python, etc.
LLMs can also produce diverse outputs because the temperature parameter controls the balance between creativity and predictability when sampling.
%
\polytest's novel idea is to systematically leverage these properties in two complementary ways: \emph{(1) Cross-lingual test generation}: Tests are generated in $n$ multiple languages at temperature zero and then unified; \emph{(2) Diverse test sampling}: $n$ Multiple test sets are generated within the same language at a higher temperature before unification. 
A key idea is that LLMs may produce tests with the same input but conflicting expected outputs across languages and generations -- a problematic inconsistency.
\polytest addresses the challenge of contradictory tests by unifying outputs generated across different languages and multiple generations. Rather than depending on a single test output, \polytest samples multiple candidate tests and reconciles them—ensuring self-consistency, resolving contradictions, and uncovering potential cases that a one-shot LLM generation might miss. Notably, this unification and contradiction detection process does not require on-the-fly execution of every test case. 
%
We further consider 
an additional step to amplify them, which would foster the diversity of the tests and ultimately to enhance their quality. 
We evaluate our implementation of \polytest on three LLMs, namely \emph{LLama3-70b}, \emph{GPT-4o}, and \emph{GPT-3.5}. We use the latest version of the popular dataset of  EvalPlus\footnote{https://github.com/evalplus/evalplus} \cite{liu2024your} that contains curated prompts of coding problems and canonical solutions (i.e., reference code). 
We evaluate our approach on 164 problems at a temperature = 0. For each problem, we generate and amplify tests in four programming languages (Java, C, Python, and JavaScript) and in a language-agnostic input/output format (CSV). We then, performed the same five times per individual language at temperature = 1, hence, covering both setups of \polytest. We perform the union of the tests at each step with our \polytest approach, i.e., between the 5 languages and between the 5 generations per language.
%
We observe up to \emph{\textbf{15.42\%}} contradicting tests (in JavaScript with GPT-4o), \emph{\textbf{7.41\%}} (in C with Llama3-70b), and \emph{\textbf{6.51\%}} (in Java with GPT-3.5). 
Results also  show that \polytest is indeed able to improve the obtained tests w.r.t. all metrics we considered, namely number of tests and passing tests (multiplied up to \emph{\textbf{x2.67}} and \emph{\textbf{x2.85}}), statement and branch coverage (improved up to \emph{\textbf{+7.9\%}} and \emph{\textbf{+9.01\%}}), and mutation score (improved up to \emph{\textbf{+11.23\%}}). Ultimately, \polytest outperforms 
single-language and single-attempt approaches without requiring continuous executions of test cases. \polytest can be particularly beneficial for programming languages where LLMs exhibiting weak performance and low test quality. 
Finally, \polytest also outperformed Pynguin, as a baseline comparison, in generated/passing tests and mutation score, with an equivalent coverage.
 

To summarize, our main contributions are as follow:

\begin{enumerate}
    \item A Novel approach \polytest to enhance tests generation with. To the best of our knowledge, it is the first automatic approach taking advantage of the diversity induced by the polyglot feature and temperature-control of LLMs. 

    \item We report a qualitative analysis of how equivalent, different, and contradicting obtained tests are between different languages and between multiple generations per language. 
    
    \item Empirical evaluation of \polytest and comparison with five languages, five generations, and to Pynguin as a baseline, showing gains and best performance with \polytest for both steps of generation and amplification.   

    \item Publicly available implementation and results on the EvalPlus benchmark \cite{replicationpackage} for reproducibility. 
    
\end{enumerate}

\section{Motivating example}



\begin{figure}[t]
\vspace{-0.5em}
\centering
\includegraphics[width=0.4\textwidth]{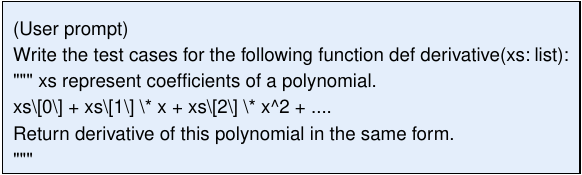}
\vspace{-0.5em}
\caption{A prompt for the derivative of a polynomial.}
\label{fig:prompt-example}
\vspace{-2em}
\end{figure}

\definecolor{squareGreen}{HTML}{eaf9db}
\definecolor{squareOrange}{HTML}{f4e7cb}

\begin{figure}[t]
\vspace{-0.5em}
\centering
\includegraphics[width=0.4\textwidth]{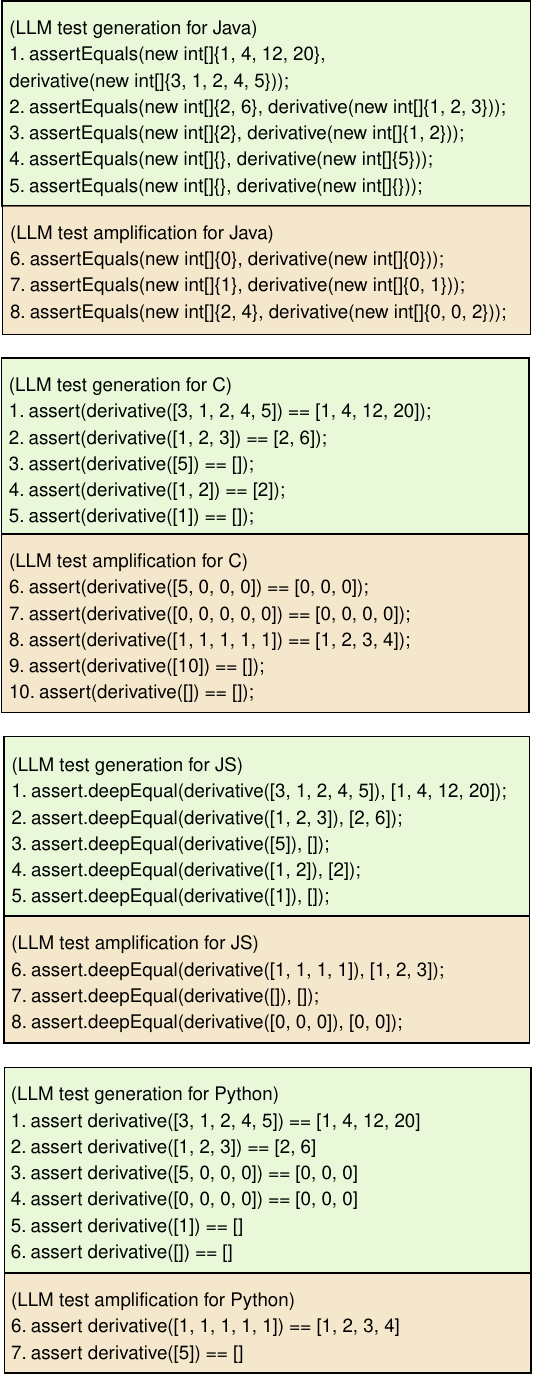}
\vspace{-0.5em}
\caption{LLM \colorbox{squareGreen}{generated} and \colorbox{squareOrange}{amplified} tests for Java, JS, C, and Python.}
\label{fig:prompt-results}
\vspace{-1.5em}
\end{figure}

\begin{figure*}[t]
\centering
\includegraphics[width=0.8\textwidth]{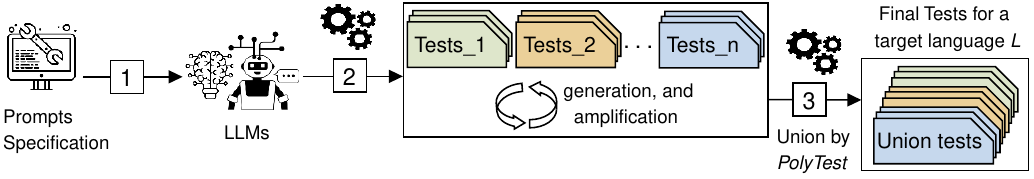}
\vspace{-1em}
\caption{Overall approach of \polytest. It covers two setups: 1) One generation of tests for \emph{n} languages and 2) \emph{n} generations for a single language. It also include three steps, generation, amplification, and reduction of tests. }
\label{fig:appraoch}
\vspace{-1em}
\end{figure*}

This section introduces a motivating example to illustrate the test generation for multiple languages and the effect of their unification. 

Let us take as an example a prompt specification for the problem of computing the derivative of a polynomial, as shown in Figure \ref{fig:prompt-example}. 
Figure \ref{fig:prompt-results} shows the results of test generation at temperature zero for four target languages, namely for Java, Javascript (JS), C, and Python. They are in the format of an \emph{assert} with an \emph{input} for the derivative function and an expected \emph{output}. 

We first observe that the LLM generates similar tests when prompted with the same problem to solve in multiple languages. However, it does generate different tests in some cases. For example, the last test $n^o$5 in Java is not present in JS and C, and vice versa, while both tests are present in Python. In addition, tests $n^o$3 and $n^o$4 in Python are not generated in other languages. This gives us a solid hint that indeed LLMs may generate different tests depending on the target language. 
Moreover, when asking the LLM to amplify the tests, we start to observe real divergences, i.e., tests with different inputs and outputs. For example, the test $n^o$8 in Java is not proposed in any of the other four languages. Similarly for the test $n^o$6 in JS and the tests $n^o$9 in C. 
As generated tests can differ for the same prompt problem on different languages, it is a diversity to take advantage of to explore unification for enhancing the test suite and various quality metrics, such as coverage and mutation score. 
Furthermore, this is likely also true when multiple generations for the same language is performed with a high temperature to allow for more creativity and diversity. For example, generating four times at temperature 1 in Java rather than one time in Java, JS, C, and Python. Ultimately, to unify the generated tests and enhance the quality of the test suite. 

However, to the best of our knowledge there is no approach allowing to automatically leverage this diversity of generated tests based on the LLM polyglot and temperature, and more importantly no empirical evaluation exists on how much unifying LLM's generated tests improves the test suite. We fill this gap in this work. 


\section{Approach}
\label{approach}


This section introduces our approach \polytest. 
The rationale and vision behind it is to reach a consensus through self-consistency for test suite and its quality. A given LLM can be weak in testing one language or in one shot iteration, but strong in testing another language or another iteration. This strength induced by the LLMs diversity can be unified to be capitalized on to improve the tests. This can be seen as a self-consistent approach.  

Figure \ref{fig:appraoch} shows the overall approach and its workflow. 
The first step {\small\boxed{1}} is to use a given LLM to then generate tests for a set of prompts {\small\boxed{2}}. Here two setups are covered, namely: 1) one generation of tests for \emph{n} different languages and 2) \emph{n} generations for a single language.
After that, \polytest performs a union of the different sets of tests {\small\boxed{3}} by aggregating the different tests and translating them in one chosen target language. 
In this step, \polytest removes duplicates and only keeps a unique occurrence of each test. 

One particularity of \polytest is its treatment of the tests in two distinct steps. First, it asks a given LLM to generate tests from the prompts. Then, \polytest asks the LLM to amplify the tests, i.e., to generate even more relevant tests. Herein, the amplification is done on top of the generated ones.  
The rationale behind the amplification step is to let the LLM propose other tests likely covering different inputs and scenarios.
%
\polytest collects the different sets of tests at each step. Hence, we can compare their quality later on. However, the developers using \polytest can freely use the results of one of the two steps w.r.t. their needs. 


Algorithms \ref{algo:setup1} and \ref{algo:setup2} details how \polytest works for the two setups, respectively, one generation of tests for \emph{n} different languages and \emph{n} generations for a single language. 
In Algorithm \ref{algo:setup1}, given a prompt $p$ specifying a given problem, 
an $LLM$, a target language and a list of languages (lines 0-2), \polytest will generate and amplify the tests for all the $l$ different languages (Lines 4-11). It
first requests the LLM to generate a set of tests for $p$ in each language $l$ (Line 5). 
Then, it will request to amplify the tests (Line 6). It does so by ending the prompt with respectively, \emph{"Generate unit tests."} and \emph{"Amplify the provided unit tests."}. 
The tests results for each step are stored for each language (Lines 7-8). 
Similarly, in Algorithm \ref{algo:setup2}, given a prompt $p$ specifying a given problem, 
an $LLM$, a target language and a number of generations $NbrGen$ (lines 0-2), \polytest will generate and amplify the tests $NbrGen$ times for the target language $TL$ (Lines 4-11).
After that, for each of the two steps, \polytest perform in Algorithm \ref{algo:union} the union of the multiple sets of tests (Lines 3-8). 
It then converts the tests from the different languages into one chosen target language $TL$. The conversation is done through the LLM (Lines 5-7). 
Tests are added once in the final unified set of tests, hence, ignoring duplicates. 
Ultimately, \polytest resulting also in unified tests corresponding to the two steps of generation and amplification.




\definecolor{circlegreen}{HTML}{7ed321}

\setlength{\textfloatsep}{5pt}

\begin{algorithm2e}[t]
 \small
\SetAlgoLined
\KwData{Prompt, LLM, Target Laguage TL, Languages}
p $\leftarrow$ Prompt

test\_res $\leftarrow$ Languages.size() * 3 Matrix

i $\leftarrow$ 0

\For {( l $\in$ Languages)}
{
tests\_step\_1 $\leftarrow$ generationRequest(LLM, l, p) \emph{\textcolor{circlegreen}{/*prompting the LLM to generate tests*/}}

tests\_step\_2 $\leftarrow$ amplificationRequest(LLM, l, p, tests\_step\_1) \emph{\textcolor{circlegreen}{/*prompting the LLM to amplify the generated tests*/}}


test\_res[i, 0].add(tests\_step\_1) \emph{\textcolor{circlegreen}{/*storing the tests of the two steps per language*/}}

test\_res[i, 1].add(tests\_step\_2)


i++
}

unionTests $\leftarrow$ UnionTests(test\_res, LLM, TL)  \emph{\textcolor{circlegreen}{/*Algorithm \ref{algo:union}*/}}

 \caption{\polytest in first scenario of one generation in multiple languages.} 
 \label{algo:setup1}

\end{algorithm2e}

\begin{algorithm2e}[t]
 \small
\SetAlgoLined
\KwData{Prompt, LLM, Target Laguage TL, number of generation NbrGen}
p $\leftarrow$ Prompt

test\_res $\leftarrow$ NbrGen * 3 Matrix

\For {( i $\in$ NbrGen)}
{
tests\_step\_1 $\leftarrow$ generationRequest(LLM, TL, p) \emph{\textcolor{circlegreen}{/*prompting the LLM to generate tests*/}}

tests\_step\_2 $\leftarrow$ amplificationRequest(LLM, TL, p, tests\_step\_1) \emph{\textcolor{circlegreen}{/*prompting the LLM to amplify the generated tests*/}}


test\_res[i, 0].add(tests\_step\_1) \emph{\textcolor{circlegreen}{/*storing the tests of the two steps per generation*/}}

test\_res[i, 1].add(tests\_step\_2)


i++
}

unionTests $\leftarrow$ UnionTests(test\_res, LLM, TL)  \emph{\textcolor{circlegreen}{/*Algorithm \ref{algo:union}*/}}

 \caption{\polytest in second scenario of multiple generations in one single language.} 
 \label{algo:setup2}
\end{algorithm2e}

\begin{algorithm2e}[t]
 \small
\SetAlgoLined
\KwData{tests\_matrix, LLM, Target Laguage TL}

tests $\leftarrow$ tests\_matrix

unionTests $\leftarrow$ \{$\phi$\} \emph{\textcolor{circlegreen}{/*a set with non-duplicate elements*/}}

\For {( t $\in$ tests)} 
{\emph{\textcolor{circlegreen}{/*unify the tests for each of the generated and amplified tests for all languages or generations*/}}

unionTests[0] $\leftarrow$ unionTests[0] $\cup$ convertTest(LLM, TL, t[0])

unionTests[1] $\leftarrow$ unionTests[1] $\cup$ convertTest(LLM, TL, t[1])


}

\textbf{return} unionTests

 \caption{UnionTests() 
 \\ Union of generated, amplified, and reduced tests.}
 \label{algo:union}
\end{algorithm2e}

\section{Methodology}

This section describes our empirical evaluation of \polytest and whether leveraging the diversity of LLMs through polyglot feature and temperature change would yield better results for test generation. 
The section first presents the selected LLMs, then our dataset, research questions, and finally the evaluation process. 

\vspace*{-2em}
\subsection{Selected LLMs and parameterization}\label{selectedLLM}

We chose 
\emph{llama3-70b}, \emph{GPT-4o}, and \emph{GPT-3.5}. 
These three models are popular with with good performances that are accessible for us and available with no or a small cost. Thus, easing future replication and reproduction of our results. 
%
The \emph{temperature hyperparameter} is usually suggested to be set between~0 and~1 in the documentation. 
The lower the temperature, the more deterministic the results are (e.g., at 0). 
Increasing temperature could lead to more diversity, creativity, and randomness. 
Therefore, we set the temperature of the LLMs to different values for the two setups of \polytest. When generating tests once for $n$ different languages, we set the temperature to zero (0). Thus, we only leverage the diversity brought by the multiple polyglot languages.  
When generating $n$ times tests for a single language, we set the temperature to one (1), hence, leveraging only on the diversity brought by the high temperature.



\subsection{Dataset}

This section details our selected dataset. We chose EvalPlus\footnote{https://github.com/evalplus/evalplus} \cite{liu2024your}, the lastest up-to-date dataset that builds on top of two existing benchmarks, namely HumanEval\footnote{https://github.com/openai/human-eval} \cite{chen2021evaluating} and MBPP\footnote{https://github.com/google-research/google-research/tree/master/mbpp} \cite{austin2021program}. 
EvalPlus benchmark is a dataset designed to evaluate the code generation capabilities of large language models (LLMs). 
It has 164 problems consisting of hand-crafted programming problems, each including a function signature, docstring, and body of canonical solution (i.e., reference code) that is important in our evaluation to verify the correctness of the obtained tests. 

\subsection{Research Questions}

This section presents the research questions for our empirical study and evaluation of \polytest for test generation w.r.t. single general purpose languages (GPLs). 

\begin{itemize}
    \item[RQ1] Does \polytest increases the number of obtained tests? This aims to quantify the improvements or not in terms of number of tests.  
    
    \item[RQ2] How much obtained tests are equivalent, different, and in contradiction across the different GPLs? 
    This aims to assess how the obtained tests in different languages or generations are alike, different, and even contradicting. 
    
    \item[RQ3] How do obtained tests that pass or fail vary across GPLs and \polytest? 
    This aims to check the correctness of the obtained tests before and after being unified with \polytest. Note that failing tests are supposedly wrong, since we have the canonical solutions to check this in our dataset. 
    
    \item[RQ4] How much statement coverage and branch coverage vary across GPLs and \polytest?
    This aims to check the quality of the obtained tests before and after being unified with \polytest with the coverage metric. 
    
    \item[RQ5] How does mutation score vary across GPL and \polytest? 
    This aims to check a crucial quality metric of the obtained tests before and after being unified with \polytest. 
    
    \item[RQ6] How does \polytest compare to a baseline of test generation? This positions \polytest with SOTA Pynguin tool \cite{lukasczyk2023empirical}.  
    
    
    
\end{itemize}

\subsection{Evaluation Process}

We launched \polytest for each of the prompts in our dataset by using the APIs of our two LLMs. We chose to include in \polytest the following languages: Java, C, Python, JavaScript, and also to ask for language-agnostic tests in the form of input/output in a CSV format. We then chose Python as a target language for unification of tests, but it could have been any other language. 
In fact the target language does not change the results of \polytest, since the unification algorithms will not be impacted and remains the same.
Note that the tests for the other languages are translated with the LLM to Python to execute them on the canonical solutions. We later on checked a random subset and confirmed the correctness of the translation (see section \ref{threats}). 
We store all intermediate and final results, i.e., tests per language, unified tests, for generation and amplification steps. 
Thus, we could later on compare them, check their correctness and quality by computing various quality metrics. In particular, \emph{statement + branch coverage} and \emph{mutation score}. We reuse coverage.py\footnote{https://coverage.readthedocs.io/en/7.5.0/} and mut.py\footnote{https://github.com/mutpy/mutpy} tools to compute the coverage and mutation metrics. They are computed as follows:


$statement \ coverage =  \dfrac{nbr \ of \ visited \ Statements}{nbr \ of \ total\ Statements} \times 100$


$branch \ coverage = \dfrac{nbr \ of \ visited \ branches}{nbr \ of \ total \ branches} \times 100$


$mutation \ score = \dfrac{nbr \ of \ killed \ mutants}{nbr \ of \ total \ mutants} \times 100$

Our dataset and implementation are publicly available in \cite{replicationpackage}.

\sloppy 

\section{Results}


As we evaluate on Java, C, Python, JS, and CSV, in the remaining of the results section, we will refer to obtained tests with \polytest from one generation for $5$ different languages as $\polytest_{5\_lang}$ and form $5$ generations per language as $\polytest_{Java\times5}$, $\polytest_{C\times5}$, $\polytest_{Python\times5}$, $\polytest_{JS\times5}$, and $\polytest_{CSV\times5}$.

\subsection{RQ1}

First we ran our experimental protocol to obtain the tests to unify with \polytest in the different setups. 

Column 4 in Table \ref{table:resultsllama} for \emph{llama3-70b}, Table \ref{table:resultsGPT4} for \emph{GPT-4o}, and Table \ref{table:resultsGPT3.5} for \emph{GPT-3.5} gives the total number of obtained test for the generation and amplification steps of our approach. 
We observe that \polytest outperforms other single language generations in number of total tests. For \emph{llama3-70b}, \polytest (in all six setups) multiples, on average, the generated and amplified tests by respectively x2.16 and x2.67. 
For \emph{GPT-4o}, \polytest multiples, on average, the generated and amplified tests by respectively x2.16 and x2.67. 
For \emph{GPT-3.5}, \polytest multiples, on average, the generated and amplified tests by respectively x1.8 and x2.3. 
\polytest thus increases significantly the number of tests compared to each single language.


\begin{tcolorbox}[boxsep=-2pt]
\textbf{$\boldsymbol{RQ_1}$ insights:}
All \polytest setups allows to increase the number of generated and amplified tests, respectively, by \emph{\textbf{x2.16}} and \emph{\textbf{x2.67}} for \emph{llama3-70b}, by \emph{\textbf{x2.16}} and \emph{\textbf{x2.67}} for \emph{GPT-4o}, and by \emph{\textbf{x1.8}} and \emph{\textbf{x2.3}} for \emph{GPT-3.5}. 
This is non-negligible gains. 
\end{tcolorbox}

\subsection{RQ2}

To answer this question, we compared the obtained tests between the different languages for $\polytest_{5\_lang}$ and between the five generations per language for $\polytest_{Java\times5}$, $\polytest_{C\times5}$, $\polytest_{Python\times5}$, $\polytest_{JS\times5}$, and $\polytest_{CSV\times5}$. In particular, we compared them pairwise two by two. 


First, we observe that great number of tests are equivalent and do overlap between the different pairs of languages and generations per language. The overlap is observed more with $\polytest_{5\_lang}$ than with the other setups. Due to lack of space the figures illustrating the overlap percentages (\%) are in our data set accessible online \cite{replicationpackage}. 
For \emph{llama3-70b}, the overlap varies from 3.62\% up to 78.7\% in the generated tests and from 4.71\% up to 62.5\% in the amplified tests. 
For \emph{GPT-4o}, the overlap varies from 4.11\% up to 80.5\% in the generated tests and from 5.17\% up to 53\% in the amplified tests. 
For \emph{GPT-3.5}, the overlap is more present in the generation step compared to the amplification step, especially with $\polytest_{5\_lang}$. The overlap varies from 1\% up to 94.5\% in the generated tests and from 1\% up to 64.2\% in the amplified tests. 
Still, an important part of the generated and amplified tests overall differed between the different languages and generations. 



\begin{figure}[t]
\vspace{-0.5em}
\centering
\hspace*{-0.85cm}
\includegraphics[width=0.53\textwidth]{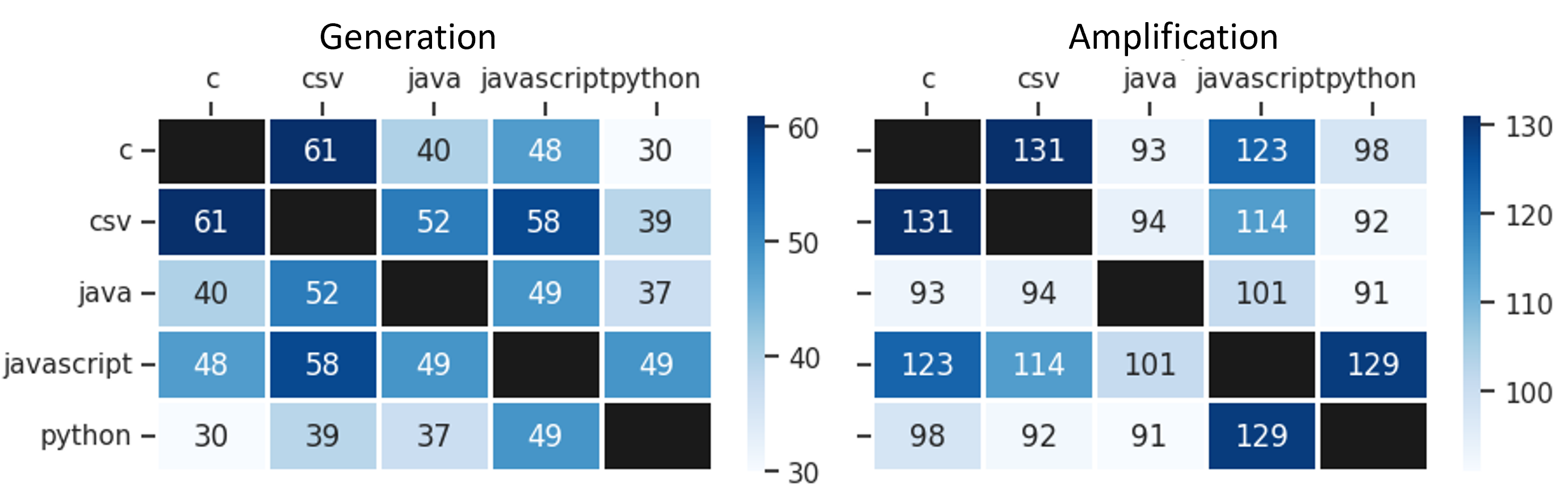}
\vspace{-0.5em}
\caption{Contradiction amid language pairs in \emph{llama3-70b}.}
\label{fig:RQ1.1}
\vspace{-1em}
\end{figure}


\begin{figure}[t]
\centering
\hspace*{-0.85cm}
\includegraphics[width=0.53\textwidth]{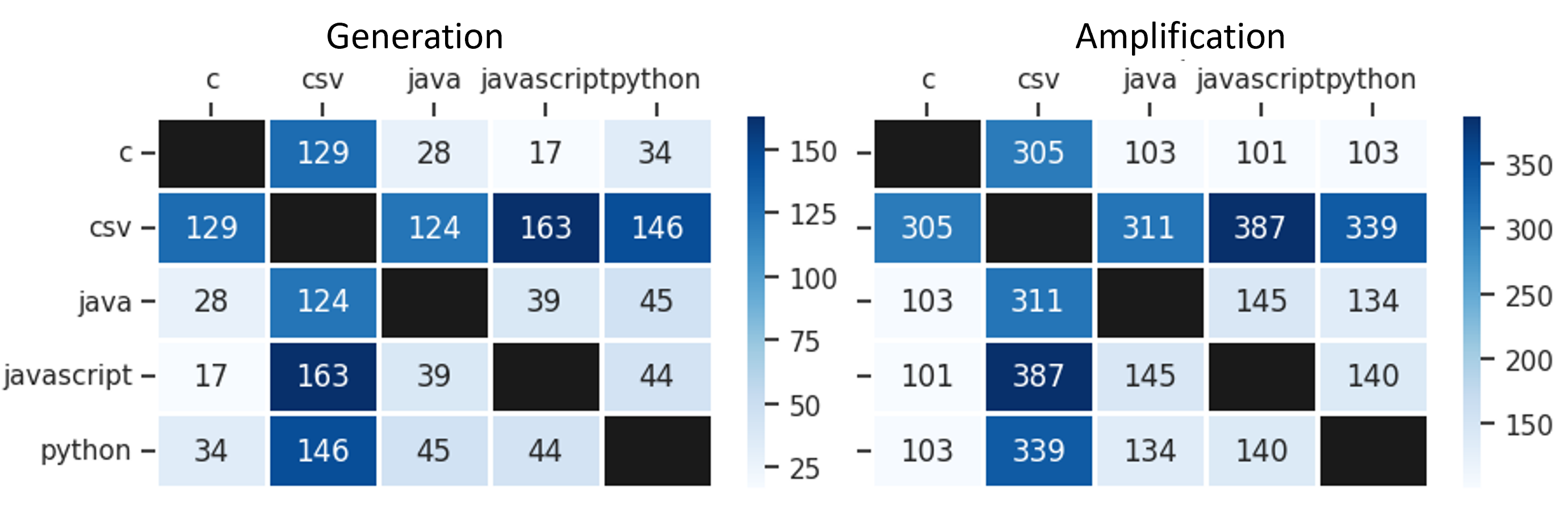}
\vspace{-0.5em}
\caption{Contradiction amid language pairs in \emph{GPT-4o}.}
\label{fig:RQ1.2}
\vspace{-1em}
\end{figure}

\begin{figure}[t]
\centering
\hspace*{-0.85cm}
\includegraphics[width=0.53\textwidth]{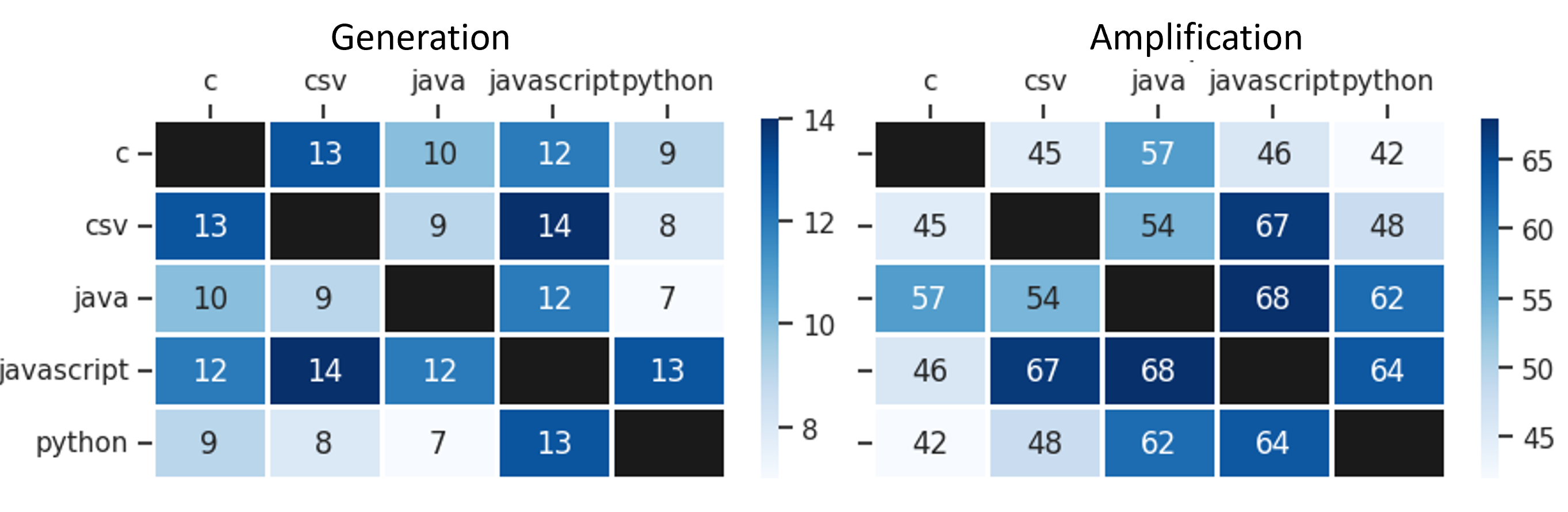}
\vspace{-0.5em}
\caption{Contradiction amid language pairs in \emph{GPT-3.5}.}
\label{fig:RQ1.3}
\vspace{-0.5em}
\end{figure}



\begin{figure}[t]
\vspace{-0.5em}
\centering
\includegraphics[width=0.45\textwidth]{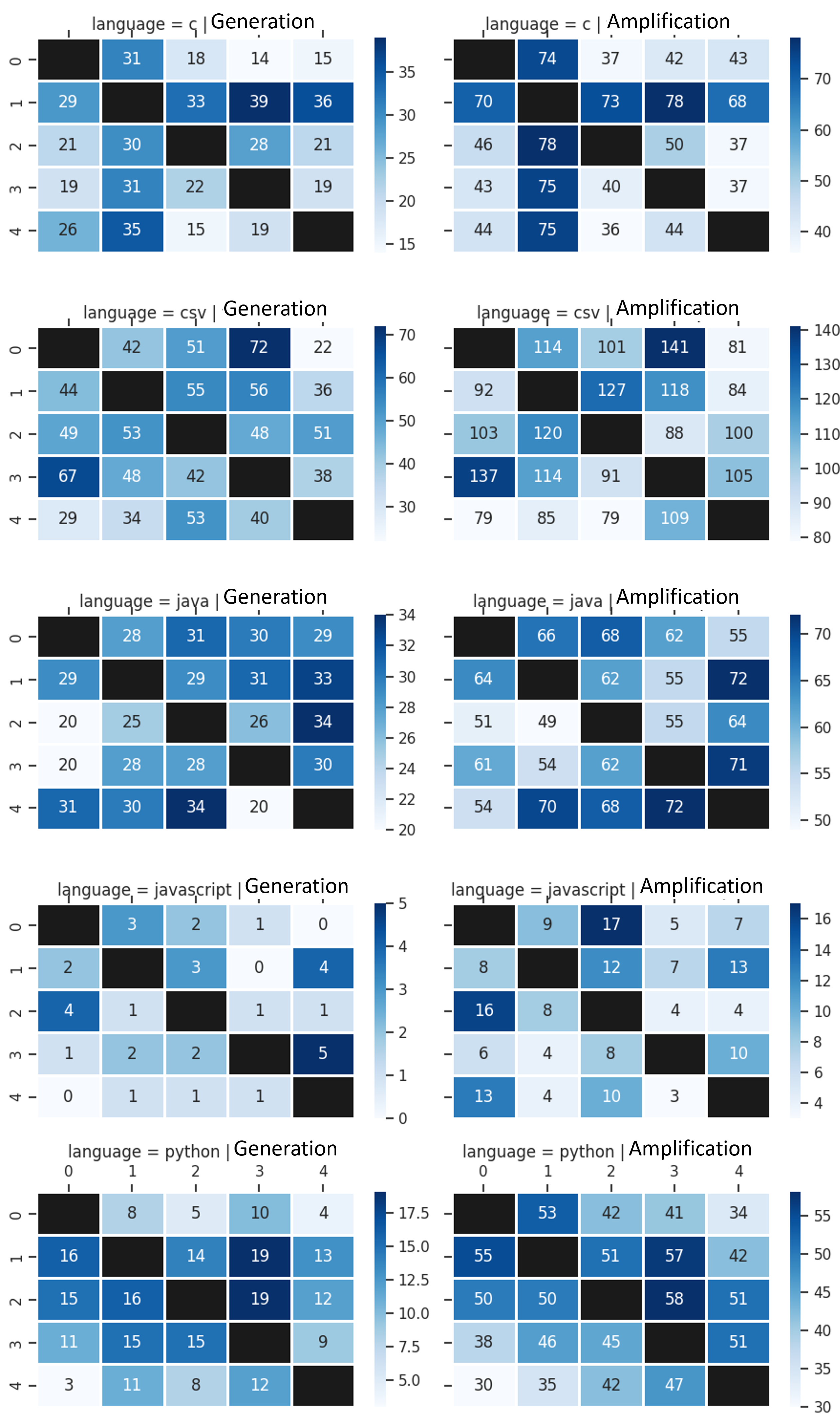}
\vspace{-0.5em}
\caption{Contradiction amid pairs in the five generations per language for \emph{llama3-70b}.}
\label{fig:RQ1.4}
\end{figure}

\begin{figure}[t]
\vspace{-0.5em}
\centering
\includegraphics[width=0.45\textwidth]{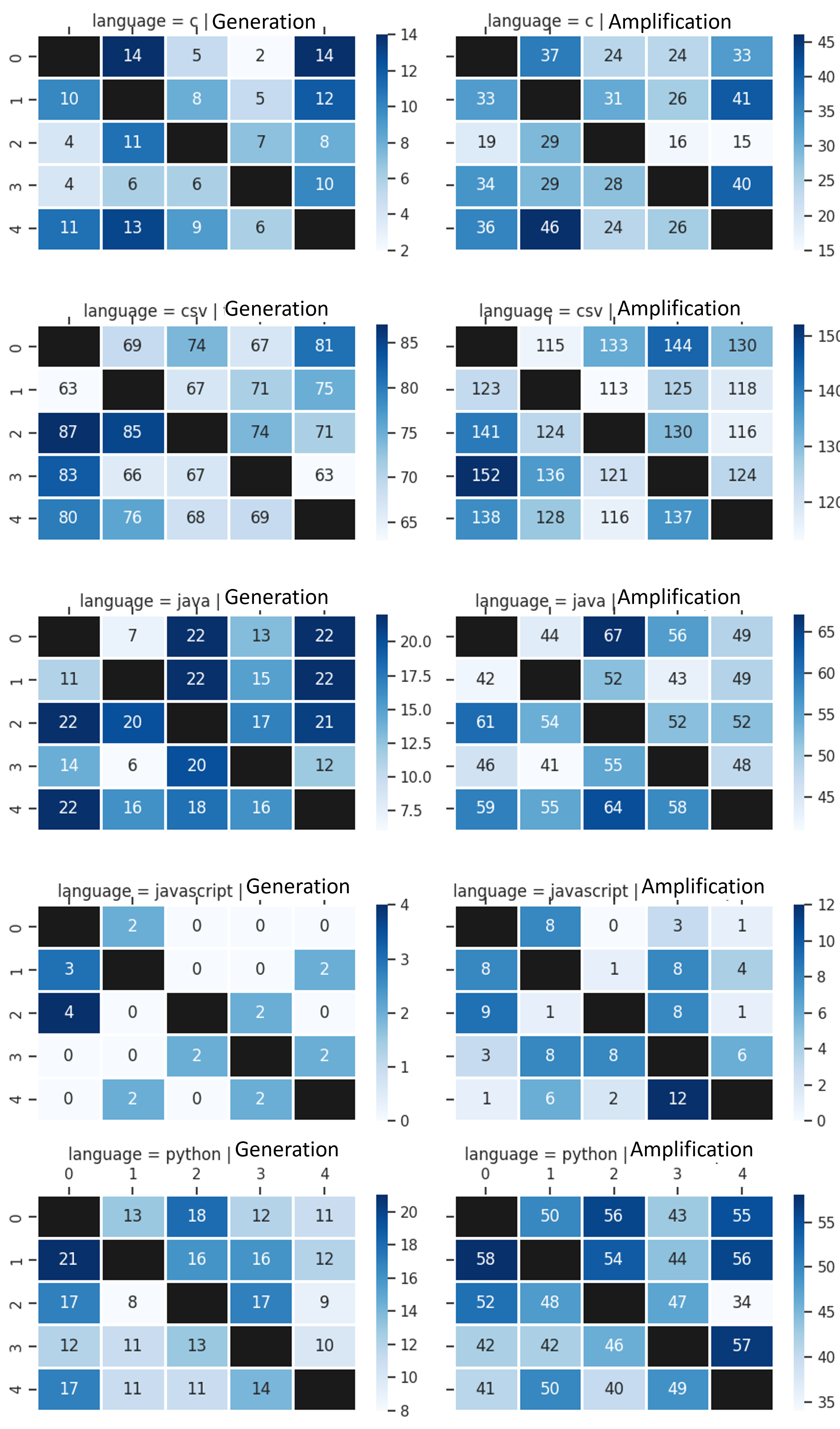}
\vspace{-0.5em}
\caption{Contradiction amid pairs in the five generations per language for \emph{GPT-4o}.}
\label{fig:RQ1.5}
\end{figure}

We further looked at the contradictions between the tests in different languages and generations. 
Our hypothesis is that generated tests for the same prompt problem are not contradicting themselves. Meaning that if they have the same input, they should have the same output as well. 
To verify our hypothesis, we searched for tests sharing the same input but have different outputs. 
Figures \ref{fig:RQ1.1}, \ref{fig:RQ1.2}, \ref{fig:RQ1.3} show the number of contradicting tests between each pair of languages and Figures \ref{fig:RQ1.4}, \ref{fig:RQ1.5} show the number of contradicting tests between the different generations per language for \emph{llama3-70b} and \emph{GPT-4o} (we ommit the Figure for \emph{GPT-3.5} due to lack of space). 
We first observe that the three LLMs do generate contradicting tests in both generation and amplification steps, which rejects our hypothesis. 
We observe slightly more contradicting tests with $\polytest_{5\_lang}$ 
than with $\polytest_{Java\times5}$, $\polytest_{C\times5}$, $\polytest_{Python\times5}$, $\polytest_{JS\times5}$, and $\polytest_{CSV\times5}$. 

While these cases still represent only a small part of all the obtained tests, we nonetheless observe them between almost all pairs of languages or generations. In particular, with a maximum of contradicting tests in the amplified step up to 15.42\% in JavaScript with \emph{GPT-4o}, 7.41\% in C with \emph{Llama3-70b}, and 6.51\% in Java with \emph{GPT-3.5}.  
Overall, the maximums of contradicting tests in the different \polytest setups were as follows. 
For \emph{llama3-70b},  131 in C with $\polytest_{5\_lang}$, 78 in C with $\polytest_{C\times5}$, 141 in CSV with $\polytest_{CSV\times5}$, 72 in Java with $\polytest_{Java\times5}$, 17 in Javascript with $\polytest_{JS\times5}$, and 58 in Python with $\polytest_{Python\times5}$. 
For \emph{GPT-4o}, the maximum of contradicting tests was 387 in Javascript with $\polytest_{5\_lang}$, 46 in C with $\polytest_{C\times5}$, 152 in CSV with $\polytest_{CSV\times5}$, 67 in Java with $\polytest_{Java\times5}$, 12 in Javascript with $\polytest_{JS\times5}$, and 58 in Python with $\polytest_{Python\times5}$. 
For \emph{GPT-3.5}, the maximum of contradicting tests was 68 in Java with $\polytest_{5\_lang}$, 36 in C with $\polytest_{C\times5}$, 25 in CSV with $\polytest_{CSV\times5}$, 21 in Java with $\polytest_{Java\times5}$, 3 in Javascript with $\polytest_{JS\times5}$, and 27 in Python with $\polytest_{Python\times5}$. 
 
These cases of contradicting tests emphasize the need to verify and validate the correctness of the tests. 
\polytest can be seen as a kind of self-consistency validation since it will consider the different contradicting tests between the different languages. Hence, \polytest will likely keep the correct passing ones at the end, which could not be the case for a single language or a single generation. In our case, since we have the canonical solutions, we can filter the wrong contradictions that will make the tests fail from those that pass. We further look into this in the next RQ. 

\begin{tcolorbox}[boxsep=-2pt]
\textbf{$\boldsymbol{RQ_2}$ insights:}
A great number of tests are in common between the different languages and generations, yet with a flagrant diversity. 
Surprisingly, contradicting tests were systematically observed in between almost all pairs of languages and the multiple generations per language in all three LLMs. The maximum of contradicting tests were up to \emph{\textbf{15.42\%}} with \emph{GPT-4o}, \emph{\textbf{7.41\%}} with \emph{Llama3-70b}, and \emph{\textbf{6.51\%}} with \emph{GPT-3.5}.  

\end{tcolorbox}

\begin{table*}
        
        \caption{Results for \polytest with llama3-70b.} 
        \label{table:resultsllama}
        \resizebox{0.7\textwidth}{!} {
        \begin{tabular}{lll cc ccc}
        \toprule
        \centering
        Temperature & Language & Step & \begin{tabular}[c]{@{}l@{}}$n^o$ of total \\ test \end{tabular}& \begin{tabular}[c]{@{}l@{}}$n^o$ of passing \\ tests\end{tabular} & \begin{tabular}[c]{@{}l@{}}Statement \\ coverage \end{tabular}& \begin{tabular}[c]{@{}l@{}}Branch \\ coverage\end{tabular} & \begin{tabular}[c]{@{}l@{}}Mutation \\ score\end{tabular} \\ \midrule

        &   & Gen. & 983 & 814  &   94.71\% & 93.40\% & 83.73\%  \\ 
        &  C & Ampl. &  1769 & 1291  &  92.18\% & 90.42\% & 82.74\%  \\ 
       \cmidrule{2-8}

         &  & Gen. &  1112 & 912 &  96.16\% & 95.19\% & 87.66\%  \\ 
        &  CSV & Ampl. & 2336 & 1774 &   96.45\% & 95.42\% & 88.63\%  \\ 
        \cmidrule{2-8}

          &   & Gen. & 843 & 737 &  98.44\% & 97.32\% & 88.70\%  \\ 
       \emph{temp=0} &  Java & Ampl. &  1487 & 1181 &  98.23\% & 97.29\% & 88.34\%  \\ 
       \cmidrule{2-8}

         &   & Gen. &  973 & 831 &  97.68\% & 96.65\% & 87.93\%  \\ 
        &  Javascript & Ampl. &  1745 & 1379 &   97.55\% & 96.50\% & 88.46\%  \\ 
        \cmidrule{2-8}

       &  & Gen. &  994 & 858 &  98.90\% & 98.09\% & 89.21\%  \\ 
        &  Python & Ampl. & 1843 & 1455 &  98.85\% & 97.94\% & 90.09\%  \\ 
       \cmidrule{2-8}


         &  & Gen. & 2180 \cellcolor{green!35}& 1634 \cellcolor{green!55}&   99.05\% \cellcolor{green!55}&  98.34\% \cellcolor{green!55}&  91.71\% \cellcolor{green!65} \\ 
       &   $\polytest_{5\_lang}$ & Ampl. & 5179 \cellcolor{green!55}& 3543 \cellcolor{green!55}&  98.94\% \cellcolor{green!55}& 97.87\% \cellcolor{green!35}&  93.53\% \cellcolor{green!75} \\
       \midrule \midrule


        &   & Gen. & 2207 \cellcolor{green!55}& 1496  \cellcolor{green!35}&   98.49\% \cellcolor{green!35}& 97.71\% \cellcolor{green!35}& 88.92\% \cellcolor{green!35} \\ 
        &  $\polytest_{C\times5}$ & Ampl. & 4714  \cellcolor{green!35}& 2999  \cellcolor{green!35}&  98.91\% \cellcolor{green!55}& 98.23\% \cellcolor{green!55}& 91.40\% \cellcolor{green!35} \\ 
       \cmidrule{2-8}

         &  & Gen. &  2636 \cellcolor{green!75}& 1884 \cellcolor{green!75}& 97.97\% \cellcolor{green!35}& 97.38\% \cellcolor{green!35}& 90.48\% \cellcolor{green!35} \\ 
        &  $\polytest_{CSV\times5}$ & Ampl. & 5999 \cellcolor{green!75}& 3814 \cellcolor{green!75}&   97.69\% \cellcolor{green!35}& 97.03\% \cellcolor{green!35}& 90.43\% \cellcolor{green!35} \\ 
       \cmidrule{2-8}

          &   & Gen. & 1833 \cellcolor{green!35}& 1337 \cellcolor{green!35}&  99.06\% \cellcolor{green!55}& 98.43\% \cellcolor{green!55}& 90.85\%\cellcolor{green!35}  \\ 
       \emph{temp=1} &  $\polytest_{Java\times5}$ & Ampl. &  4127 \cellcolor{green!35}& 2722 \cellcolor{green!35}&  99.38\% \cellcolor{green!75}& 98.83\% \cellcolor{green!75}& 92.08\% \cellcolor{green!35} \\ 
       \cmidrule{2-8}

         &   & Gen. &  2092 \cellcolor{green!35}& 1576 \cellcolor{green!35}&  99.28\% \cellcolor{green!75}& 98.55\% \cellcolor{green!55}& 91.51\% \cellcolor{green!55} \\ 
        &  $\polytest_{JS\times5}$ & Ampl. &  4652 \cellcolor{green!35}& 3180 \cellcolor{green!35}&   99.46\% \cellcolor{green!75}& 98.85\% \cellcolor{green!75}& 92.52\% \cellcolor{green!55} \\ 
       \cmidrule{2-8}

       &  & Gen. &  2058 \cellcolor{green!35}& 1623 \cellcolor{green!55}&  99.39\% \cellcolor{green!75}& 98.84\% \cellcolor{green!55}& 92.17\% \cellcolor{green!75} \\ 
        &  $\polytest_{Python\times5}$ & Ampl. & 4848 \cellcolor{green!35}& 3405 \cellcolor{green!55}&  99.39\% \cellcolor{green!75}& 98.92\% \cellcolor{green!75}& 93.31\% \cellcolor{green!55} \\ 


        \bottomrule 

        \end{tabular}
    }
\end{table*}

\begin{table*}
        
        \caption{Results for \polytest with GPT-4o.} 
        \label{table:resultsGPT4}
        \resizebox{0.7\textwidth}{!} {
        \begin{tabular}{lll cc ccc}
        \toprule
        \centering
        Temperature & Language & Step & \begin{tabular}[c]{@{}l@{}}$n^o$ of total \\ test \end{tabular}& \begin{tabular}[c]{@{}l@{}}$n^o$ of passing \\ tests\end{tabular} & \begin{tabular}[c]{@{}l@{}}Statement \\ coverage \end{tabular}& \begin{tabular}[c]{@{}l@{}}Branch \\ coverage\end{tabular} & \begin{tabular}[c]{@{}l@{}}Mutation \\ score\end{tabular} \\ \midrule

        &   & Gen. & 981 & 891 &   97.93\% & 97.01\% & 89.82\%  \\ 
        &  C & Ampl. &  2680 & 2281 &  98.48\% & 97.56\% & 91.02\%  \\ 
        \cmidrule{2-8}

         &  & Gen. &  1671 & 1283 &  91.63\% & 90.15\% & 83.12\%  \\ 
        &  CSV & Ampl. & 4626 & 3325 &   91.71\% & 90.25\% & 83.53\%  \\ 
        \cmidrule{2-8}

          &   & Gen. & 973 & 849 &  98.62\% & 97.63\% & 89.95\%  \\ 
       \emph{temp=0} &  Java & Ampl. &  2677 & 2166 &  98.37\% & 97.47\% & 90.83\%  \\ 
       \cmidrule{2-8}

         &   & Gen. &  1157 & 1029 &  98.20\% & 97.27\% & 90.18\%  \\ 
        &  Javascript & Ampl. &  3022 & 2510 &   96.82\% & 95.79\% & 89.22\%  \\ 
        \cmidrule{2-8}

       &  & Gen. &  1260 & 1158 &  99.09\% & 98.39\% & 91.83\%  \\ 
        &  Python & Ampl. & 3354 & 2927 &  99.04\% & 98.17\% & 93.23\%  \\ 
       \cmidrule{2-8}


         &  & Gen. & 2903 \cellcolor{green!75}& 2249  \cellcolor{green!55}&  99.48\% \cellcolor{green!75}& 98.98\% \cellcolor{green!75}& 93.33\% \cellcolor{green!55} \\ 
       &   $\polytest_{5\_lang}$ & Ampl. & 9975 \cellcolor{green!75}& 7311 \cellcolor{green!75}&  99.61\% \cellcolor{green!75}& 98.87\% \cellcolor{green!75}& 94.39\% \cellcolor{green!65} \\
       \midrule \midrule


        &   & Gen. & 1480 \cellcolor{green!35}& 1214 \cellcolor{green!35}&   99.03\% \cellcolor{green!35}& 98.32\% \cellcolor{green!35}& 90.14\% \cellcolor{green!35} \\ 
        &  $\polytest_{C\times5}$ & Ampl. &  5170 \cellcolor{green!35}& 4108 \cellcolor{green!35}&  99.54\% \cellcolor{green!65}& 99.04\% \cellcolor{green!75}& 94.45\% \cellcolor{green!65} \\ 
       \cmidrule{2-8}

         &  & Gen. &  3671\cellcolor{green!75} & 2865 \cellcolor{green!75}&  99.60\% \cellcolor{green!75}& 99.16\% \cellcolor{green!75}& 93.79\% \cellcolor{green!75} \\ 
        &  $\polytest_{CSV\times5}$ & Ampl. & 8937 \cellcolor{green!55}& 6625 \cellcolor{green!55}&   99.46\% \cellcolor{green!35}& 98.79\% \cellcolor{green!55}& 94.76\% \cellcolor{green!75} \\ 
        \cmidrule{2-8}

          &   & Gen. & 2389 \cellcolor{green!35}& 1822 \cellcolor{green!35}&  99.39\% \cellcolor{green!35}& 98.73\% \cellcolor{green!35}& 91.96\% \cellcolor{green!35} \\ 
       \emph{temp=1} &  $\polytest_{Java\times5}$ & Ampl. &  6204 \cellcolor{green!35}& 4386 \cellcolor{green!35}&  99\% \cellcolor{green!35}& 98.34\% \cellcolor{green!35}& 92.67\% \cellcolor{green!35} \\ 
       \cmidrule{2-8}

       &  & Gen. &  2657 \cellcolor{green!35}& 2167 \cellcolor{green!35}&  99.23\% \cellcolor{green!35}& 98.65\% \cellcolor{green!35}& 92.12\% \cellcolor{green!35} \\ 
        &  $\polytest_{JS\times5}$ & Ampl. & 6581 \cellcolor{green!35}& 5073 \cellcolor{green!35}&  99.04\% \cellcolor{green!35}& 98.50\% \cellcolor{green!35}& 94.07\% \cellcolor{green!35} \\ 
       \cmidrule{2-8}

         &  & Gen. &  2890 \cellcolor{green!75}& 2420 \cellcolor{green!55}& 99.43\% \cellcolor{green!75} & 98.93\% \cellcolor{green!75} & 93.30\% \cellcolor{green!55} \\ 
       &   $\polytest_{Python\times5}$ & Ampl. &  7309 \cellcolor{green!35}& 5788 \cellcolor{green!35}& 99.59\% \cellcolor{green!75} & 99.10\% \cellcolor{green!75} & 94.64\% \cellcolor{green!75} \\ 
       \bottomrule 

        \end{tabular}
    }
\end{table*}

\begin{table*}
        
        \caption{Results for \polytest with GPT-3.5.} 
        \label{table:resultsGPT3.5}
        \resizebox{0.7\textwidth}{!} {
        \begin{tabular}{lll cc ccc}
        \toprule
        \centering
        Temperature & Language & Step & \begin{tabular}[c]{@{}l@{}}$n^o$ of total \\ test \end{tabular}& \begin{tabular}[c]{@{}l@{}}$n^o$ of passing \\ tests\end{tabular} & \begin{tabular}[c]{@{}l@{}}Statement \\ coverage \end{tabular}& \begin{tabular}[c]{@{}l@{}}Branch \\ coverage\end{tabular} & \begin{tabular}[c]{@{}l@{}}Mutation \\ score\end{tabular} \\ \midrule

        &   & Gen. & 497 & 454 &   93.86\% & 91.92\% & 83.59\%  \\ 
        &  C & Ampl. &  1037 & 831 &  94.54\% & 92.80\% & 80.92\%  \\ 
        \cmidrule{2-8}

         &  & Gen. &  471 & 445 &  96.73\% & 94.70\% & 85.17\%  \\ 
        &  CSV & Ampl. & 1235 & 1028 &  98.32\% & 96.94\% & 89.97\%  \\ 
        \cmidrule{2-8}

          &   & Gen. & 485 & 457 &  96.84\% & 94.84\% & 85.25\%  \\ 
       \emph{temp=0} &  Java & Ampl. &  1045 & 802 &  97.18\% & 95.53\% & 86.34\%  \\ 
       \cmidrule{2-8}

         &   & Gen. &  516 &  477 &  95.79\% & 94.05\% & 83.89\%  \\ 
        &  Javascript & Ampl. &  1111 & 884 &   96.34\% & 94.75\% & 86.12\%  \\ 
        \cmidrule{2-8}

       &  & Gen. & 501  & 475 &  96.43\% & 94.64\% & 84.93\%  \\ 
        &  Python & Ampl. & 1134  & 906 &   97.40\% & 95.97\% & 88.78\%  \\ 
       \cmidrule{2-8}


         &  & Gen. &  659 \cellcolor{green!65}&  540 \cellcolor{green!55}&  98.47\% \cellcolor{green!65}& 96.92\% \cellcolor{green!55}& 87.35\% \cellcolor{green!65} \\ 
       &   $\polytest_{5\_lang}$ & Ampl. & 2582 \cellcolor{green!65}&  1688 \cellcolor{green!65}&  98.67\% \cellcolor{green!75}& 97.50\% \cellcolor{green!75}& 91.57\% \cellcolor{green!75} \\ 
       \midrule \midrule


        &   & Gen. & 598 \cellcolor{green!35}& 527  \cellcolor{green!35}&  96.99\% \cellcolor{green!35}& 95.16\% \cellcolor{green!35}& 84.49\% \cellcolor{green!35} \\ 
        &  $\polytest_{C\times5}$ & Ampl. & 1637 \cellcolor{green!35}& 945  \cellcolor{green!55}&  95.22\% \cellcolor{green!35}& 93.11\% \cellcolor{green!35}& 85.60\% \cellcolor{green!35} \\ 
       \cmidrule{2-8}

         &  & Gen. & 543 \cellcolor{green!35}&  471 \cellcolor{green!35}&  98.22\% \cellcolor{green!55}& 96.41\% \cellcolor{green!35}& 87.55\% \cellcolor{green!65} \\  
        &  $\polytest_{CSV\times5}$ & Ampl. & 2745 \cellcolor{green!75}& 1761  \cellcolor{green!75}&  98.41\% \cellcolor{green!55}& 96.39\% \cellcolor{green!35}& 85.99\% \cellcolor{green!35} \\ 
        \cmidrule{2-8}

          &   & Gen. & 594 \cellcolor{green!35}&  504 \cellcolor{green!35}&  98.65\% \cellcolor{green!65}& 97.22\% \cellcolor{green!75}& 88.21\% \cellcolor{green!75} \\  
       \emph{temp=1} &  $\polytest_{Java\times5}$ & Ampl. & 2276 \cellcolor{green!35}& 1214  \cellcolor{green!35}&  98.12\% \cellcolor{green!35}& 96.26\% \cellcolor{green!35}& 84.38\% \cellcolor{green!35} \\ 
       \cmidrule{2-8}

       &  & Gen. &  658 \cellcolor{green!65}&  565 \cellcolor{green!75}&  98.99\% \cellcolor{green!75}& 97.52\% \cellcolor{green!75}& 88.73\% \cellcolor{green!75} \\ 
        &  $\polytest_{JS\times5}$ & Ampl. & 2294 \cellcolor{green!35}&   1452 \cellcolor{green!35}&  98.52\% \cellcolor{green!65}& 96.74\% \cellcolor{green!35}& 89.03\% \cellcolor{green!55} \\ 
       \cmidrule{2-8}

         &  & Gen. & 706 \cellcolor{green!75}&  612 \cellcolor{green!75}&  98.52\% \cellcolor{green!65}& 97.12\% \cellcolor{green!75}& 88.55\% \cellcolor{green!75} \\  
       &   $\polytest_{Python\times5}$ & Ampl. & 2729 \cellcolor{green!75}& 1756  \cellcolor{green!75}&  98.36\% \cellcolor{green!35}& 96.74\% \cellcolor{green!35}& 89.95\% \cellcolor{green!55} \\ 
       \bottomrule 

        \end{tabular}
    }
\vspace{-1em}
\end{table*}

\subsection{RQ3}

To answer this question, we rely on the canonical solutions in our dataset that are known to be the correct implementation solution. Thus, if a test passes, we consider it correct and if it fails, we consider it incorrect. 
This way, we can check the correctness of the obtained tests for each language and after being unified with \polytest. 

Column 5 in Tables \ref{table:resultsllama}, \ref{table:resultsGPT4}, \ref{table:resultsGPT3.5} shows the number of passing tests per language and for the union with \polytest for the generated and amplified tests in the the LLMs. 
%
We observe that \polytest outperforms other single languages in number of total correct tests passing that will be kept by developers at the end. 
For \emph{llama3-70b}, \polytest (in all six setups) multiples, on average, the generated and amplified passing tests by respectively x1.92 and x2.35. 
For \emph{GPT-4o}, \polytest multiples, on average, the generated and amplified passing tests by respectively x2.62 and x2.85. 
For \emph{GPT-3.5}, \polytest multiples, on average, the generated and amplified passing tests by respectively x1.16 and x1.66. 
\polytest thus increases significantly the number of passing tests compared to each single language, especially with \emph{llama3-70b} and \emph{GPT-4o}.
While \polytest increases the passing tests compared to each single language, it does not mean that the overall quality is improved. The next RQs investigate this aspect.  

\begin{tcolorbox}[boxsep=-2pt]
\textbf{$\boldsymbol{RQ_3}$ insights:}
\polytest increased the passing tests.  
With \emph{llama3-70b} and \emph{GPT-4o}, \polytest more than double the passing tests, on average, up to \emph{\textbf{x2.35}} and \emph{\textbf{x2.85}}. 
For \emph{GPT-3.5}, \polytest multiples the passing tests, on average, by \emph{\textbf{x1.66}}.

\end{tcolorbox}

\subsection{RQ4}

To answer this question, we compute statement and branch coverage for the obtained passing tests. 

Column 6 and 7 in Tables \ref{table:resultsllama}, \ref{table:resultsGPT4}, \ref{table:resultsGPT3.5} give the average statement and branch coverage per language and all \polytest setups for the generated and amplified tests. 
Once again, for all three LLMs, 
we observe that the highest coverage metrics are obtained by \polytest (in all its setups) exceeding all other languages in both steps. 
For \emph{llama3-70b}, we observe gains in: \emph{(1)} statement coverage up to 4.68\% in the generated tests and up to 7.21\% in the amplified tests, \emph{(2)} branch coverage up to 5.44\% in the generated tests and up to 8.5\% in the amplified tests. 
For \emph{GPT-4o}, we observe gains in \emph{(1)} statement coverage up to 7.85\% in the generated tests and up to 7.9\% in the amplified tests, \emph{(2)} branch coverage up to 9.01\% in the generated tests and up to 8.85\% in the amplified tests. 
Finally, for \emph{GPT-3.5-turbo}, we observe gains in \emph{(1)} statement coverage up to 5.13\% in the generated tests and up to 4.13\% in the amplified tests, \emph{(2)} branch coverage up to 5.6\% in the generated tests and up to 4.7\% in the amplified tests.
This is an interesting result showing systematic enhanced coverage. The aggregation of the good results of some languages, such as Python, can be transferred to other less performing languages, such as C.

\begin{tcolorbox}[boxsep=-2pt]
\textbf{$\boldsymbol{RQ_4}$ insights:}
All different setups of \polytest provided gains in the statement coverage and in branch coverage. At best the gains with \polytest in the statement coverage was up to \emph{\textbf{+7.9\%}} and in branch coverage was up to \emph{\textbf{+9.01\%}}. The minimum gains were less than 0.5\% in some cases in \emph{llama3-70b} and \emph{GPT-4o}.    

\end{tcolorbox}

\subsection{RQ5}

To answer this question, we compute mutation score for the obtained passing tests. 
Column 8 in Tables \ref{table:resultsllama}, \ref{table:resultsGPT4}, \ref{table:resultsGPT3.5} gives the average mutation score per language and all \polytest setups for the generated and amplified tests. %
Once again, for all three LLMs, we observe that the highest coverage metrics are obtained by \polytest (in all its setups) exceeding all other languages in both steps.


For \emph{llama3-70b}, we observe a systematic gain in mutation score up to 8.44\% in the generated tests and up to 10.79\% in the amplified tests.
For \emph{GPT-4o}, we observe a systematic gain in mutation score up to 10.67\% in the generated tests and up to 11.23\% in the amplified tests.
Finally, for \emph{GPT-3.5}, we observe a systematic gain in mutation score up to 5.14\% in the generated tests and up to 10.65\% in the amplified tests. 
This is an important finding that highlights the combined benefits over improved mutation score from unifying tests across different languages and generations. 
Especially as the mutation score is acknowledge to be a better metric for the tests quality \cite{li2009experimental,parsai2020comparing,andrews2006using}.  


\begin{tcolorbox}[boxsep=-2pt]
\textbf{$\boldsymbol{RQ_5}$ insights:}
\polytest was able to provide a significant gain in the mutation score up to \emph{\textbf{+10.79\%}} with \emph{llama3-70b}, \emph{\textbf{+11.23\%}} with \emph{GPT-4o}, \emph{\textbf{+10.65\%}} with \emph{GPT-3.5-turbo}. 
\end{tcolorbox}


\begin{table}[h]
        \vspace{-0.5em}
        \caption{Results for Pynguin generated tests.}
        \vspace{-0.5em}
        \label{table:resultsPynguin}
        \resizebox{0.5\textwidth}{!} {
        \begin{tabular}{l cc ccc}
        \toprule
        \centering
         Algorithms & \begin{tabular}[c]{@{}l@{}}$n^o$ of total \\ test \end{tabular}& \begin{tabular}[c]{@{}l@{}}$n^o$ of passing \\ tests\end{tabular} & \begin{tabular}[c]{@{}l@{}}Statement \\ coverage \end{tabular}& \begin{tabular}[c]{@{}l@{}}Branch \\ coverage\end{tabular} & \begin{tabular}[c]{@{}l@{}}Mutation \\ score\end{tabular} \\ \midrule

        DYNAMOSA & 441 \cellcolor{red!55}& 145 \cellcolor{red!75}&   98.68\% \cellcolor{green!55}& 98.59\% \cellcolor{green!55}& 23.70\%  \cellcolor{red!55}\\  \midrule
        
        MIO & 413 \cellcolor{red!55}& 137 \cellcolor{red!75}&   98.64\% \cellcolor{green!55}& 98.62\% \cellcolor{green!55}& 16.82\%  \cellcolor{red!75}\\  \midrule 
        
        Random & 12368 \cellcolor{green!75}& 5726 \cellcolor{green!75}&   98.18\% \cellcolor{green!35}& 98.08\% \cellcolor{green!35}& 32.54\%  \cellcolor{red!65}\\  \midrule 
        
        whole-suite & 961 \cellcolor{red!55}& 107 \cellcolor{red!75}&   98.76\% \cellcolor{green!55}& 98.78\% \cellcolor{green!55}& 11.14\% \cellcolor{red!75} \\  
        
       \bottomrule 

        \end{tabular}
    }
    \vspace{-1em}
\end{table} 

\subsection{RQ6}

To position \polytest with state of the art test generation, we compare it to a baseline, namely Pynguin tool \cite{lukasczyk2023empirical} in Python and its four algorithms DYNAMOSA, MIO, Random, and whole-suite. Overall, Pynguin's results are terrible and do not compete with \polytest. 

The number of generated test and passing test is extremely low except for Random algorithm that outperforms \polytest. Only the results of statement and branch coverage are comparable to \polytest, scoring a steady 98\% as a minimum. However, on the mutation score, Pynguin shows its weakness. The mutation score varied on average from 11.14\% up to 32.54\%, which is extremely low compared to \polytest.
Finally, we observe that several of Pynguin's generated tests do not contain assertions and are unreadable compared to the \polytest ones. These results demonstrates the benefit of \polytest, in particular, w.r.t. enhancing the tests and their quality.

\begin{tcolorbox}[boxsep=-2pt]
\textbf{$\boldsymbol{RQ_6}$ insights:}
\polytest outperforms Pynguin in generated/passing tests and mutation score, except for coverage. 
\end{tcolorbox}

\subsection{Discussion of \polytest impact}

Results confirms that \polytest, in all its setups, is effective in enhancing the test suite and its quality, in particular, with the mutation score that is considered a more relevant quality metric than the coverage \cite{li2009experimental,parsai2020comparing,andrews2006using}. It also outperformed the SOTA Pynguin tool in generated Python tests.
\polytest ensures to unify the strengths of tests in each single language or each single generation to transfer it to other languages with lower performances. For example, if an LLM is not trained enough on a given language $l_{1}$ (e.g., Rust, Swift, Go, etc.) but is trained better on other languages $l_{2}, l_{3}, ... ,l_{n} $ (e.g., Python, Java, etc.). 
One can transfer with \polytest the best results for $l_{2}, l_{3}, ... ,l_{n} $ to $l_{1}$, either through $\polytest_{5\_lang}$, or through $\polytest_{C\times5}$, $\polytest_{CSV\times5}$, $\polytest_{Java\times5}$, $\polytest_{JS\times5}$, and $\polytest_{Python\times5}$. It has also the benefit of diversifying the generated an amplified tests. 
Another advantage of \polytest is its kind of self-consistency in test generation. Indeed, our results demonstrate that the three LLMs generate incorrect contradicting tests, which to the best of our knowledge no prior study has investigated or shown so far. This poses a serious issue if developers are unaware of it. \polytest can detect these contradicting tests to filter and keep the correct ones. 
In a way, instead of relying on one test output at a time, \polytest has a self-consistency by sampling multiple candidate tests and reconcile them, potentially catching contradictions/errors or omissions/uncovered cases that a single run might miss.
Therefore, \polytest benefits developers regardless their chosen programming language and can optimise further the tests quality.  

\subsection{Threats to validity and limitations}
\label{threats}

We now discuss internal and external threats to validity \cite{wohlin2012experimentation}.

\subsubsection{Internal Validity} 

As we used the EvalPlus dataset that includes canonical solutions, we do not have a threat w.r.t. testing correctness of the tests and we have a full confidence in their results, i.e., passing or failing. 
We also set temperature at zero for $\polytest_{5\_lang}$ so that our experiment and results are deterministic as much as possible, hence, increasing confidence in our results and easing reproducibility. For the other setups of \polytest, we set the temperature to 1 to leverage on its brought diversity and creativity over the 5 generations.  
We did not further vary the temperature as our goal is to investigate whether unifying tests in multiple languages or generations would improve the quality of the test suite and not the effect of the temperature on it. Thus, studying the effect of varying the temperature.

Moreover, to perform the union, \polytest transforms the tests in different languages to a target language through the LLM itself that. In our experiment the target language was Python and could have been any other language. In fact the target language does not change the results of \polytest, since the unification algorithms will not be impacted. However, the transformation of tests through LLMs raises a risk of mis-translation. To mitigate this, we run random manual checks of translated tests to check their correctness. 
We found that the translations were correct in all our random verification.

Finally, in our evaluation, we focused solely on LLM-based test generation and only compared against Pynguin \cite{lukasczyk2023empirical} since we chose Python as a target language. However, other tools exist, such as EvoSuite \cite{10.1145/2025113.2025179} for Java, and Nessie \cite{10.1145/3510003.3510106} for JavaScript. 
In contrast, \polytest supports multiple languages and can derive tests directly from specifications. 
Prior evidence suggests that LLM-based test generation can be competitive with, and in some cases even outperform, traditional tools on key metrics \cite{pan2024multi,schafer2023empirical,dakhel2024effective}. This is confirmed by our comparison with Pynguin in \emph{RQ6}. Our design study demonstrates that leveraging polyglot capabilities and temperature diversity enhances LLM-based test generation by yielding improved coverage and mutation scores. Future research should benchmark against a broader array of test generation tools across various languages to further validate and extend these findings.

\subsubsection{External Validity} 

Our approach is evaluated on three LLMs. Thus, we cannot generalize our results of \polytest on other LLMs, such as startcoder, etc. 
However, if another LLM is considered in additional to our three LLMs for the unification over multiple LLMs, at worst \polytest would not decrease the performance observed in our results and at best it would improve them. 
We further evaluated \polytest over 5 languages only. Other languages, such as Ruby, Rust, etc., could also be considered. Similarly, the results of \polytest at worst would stay the same as in Tables \ref{table:resultsllama}, \ref{table:resultsGPT4}, \ref{table:resultsGPT3.5} and at best would improve them further. 
However, we cannot generalize our results for the case of considering other languages and LLMs. This is left for future work. 
Finally, we evaluated on the dataset of EvalPlus  \cite{liu2024your} consisting of self-contained functions. Thus, we cannot generalize our results to the more complex programs with dependencies (e.g., program with several interrelated methods) that would require complex objects as input. This is a limitation of our current approach and evaluation that we plan both to adapt and enhance in future work.  
Nonetheless, by the design of the union of \polytest, the results of a given LLM on a complex program would not worsen and at best would be improved as in our results.

\section{Related Work}
\label{RelatedWork}

This section discusses close related work that focuses on empirically evaluating LLMs on test generation. 
LLMs have been applied in different domains and tasks in Software Engineering \cite{10109345,10173990,liu2023improving,hou2023large,pearce2022asleep,sobania2022choose,ziegler2022productivity,vaithilingam2022expectation,nguyen2022empirical,doderlein2022piloting,
nathalia2023artificial,yeticstiren2023evaluating,guo2023exploring,fu2023chatgpt,kabir2023empirical,chaaben2023towards,camara2023assessment,AbukhalafHK23,10344012,jiang2023selfevolve,zhang2023multilingual}. 
All the above studies focused on either evaluating the ability of LLMs to generate qualitative code, refining it, repairing it if vulnerable, or augmenting it. However, none of them specifically explored the task of test generation.  
There is only few works assessing the ability of LLMs to generate tests \cite{schafer2023empirical,siddiq2024using,baudry2024generative,sapozhnikov2024testspark,li2024large,gu2024testart}. 
However, to the best of our knowledge, they only evaluate the capability of LLMs to generate tests for a target single language and not their unification. 

Indeed, Schafer et al. \cite{schafer2023empirical} proposed TESTPILOT, an adaptive LLM-based test generation tool for JavaScript. It relies on GPT3.5-turbo at automatically generates unit tests for the methods in JavaScript projects, evaluated on 25 npm packages. It explores how different prompt components can improve the tests. 
Siddiq et al. \cite{siddiq2024using} run an empirical study on test generation to compare three LLMs, namely Codex, GPT-3.5-Turbo, and StarCoder, with a focus on test correctness. 
Baudry et al. \cite{baudry2024generative} focuses on producing fake test data and test data generators with GPT-4 for various application domains in in Chinese, Farsi, Portuguese, Sinhalese, French, Hindi, Spanish, and English. 
Sapozhnikov et al. \cite{sapozhnikov2024testspark} introduced TestSpark, a plugin for IntelliJ IDE that enables users to generate unit tests in Java. 
Li et al. \cite{li2024large} proposed a multi-agent framework called TestChain that decouples the generation of test inputs and test out
puts in Python, which gave better results than when generating the tests directly with a 13.84\% improvement.  


Lemieux et al. \cite{lemieux2023codamosa}proposed to combine
Search-based software testing with the Codex LLM to explore whether Codex can be used to help SBST’s exploration in Python. This work is interesting as it aims to improve the LLM generated test with SBST. Our work also have similar goal but follows another path by leveraging Multi-langual/Polyglot feature of LLMs. 
Nashid et al. \cite{nashid2023retrieval} proposed an approach named CEDAR that create  effective prompts
to help Codex LLM with different code-related tasks of program repair and test generation by providing example of code and tests. It was evaluated for two languages. 
Both Chen et al. \cite{chen2022codet} and Lahiri et al. \cite{lahiri2022interactive} used Codex as an LLM to generate code and test cases from problem descriptions in the prompt similarly as in this paper. 
Bareiss et al. \cite{bareiss2022code} evaluated the performance of Codex on three code-generation tasks,
including test generation. They propose embedding
contextual information into the prompt to better guide the LLM.
El Haji et al. \cite{el2024using} ran an empirical study exploring the effectiveness of GitHub Copilot at generating tests for Python. 
Gu el al. \cite{gu2024testart} proposed to improve the LLMs' generated incorrect tests with co-evolution and repair.
Pan et al. \cite{pan2024multi} conducted an empirical study to enhance the LLMs' generated tests with guidance by static analysis demonstrated on Java and Python.
Dakhel et al. \cite{dakhel2024effective} proposed to enhance the generated tests by including the surviving mutants.

However, all above approaches focus on either investigating the test generation a targeted language or aims to improves its generated tests with repair or static analysis techniques. 

Moreover, Wang et al. \cite{wang2022self} proposed a strategy of self-Consistency to improve the chain of thought in LLM, which is kind of alternative self-Consistency strategy we use in \polytest. 
%
To the best of our knowledge, no study investigated the benefit of leveraging on LLMs diversity induced from their Multi-lingual/Polyglot ability and high temperature over multiple generations to improve the quality of generated and amplified tests. We empirically evaluated how effective this novel strategy implemented in \polytest can improve the quality of the generated and amplified tests. 

\vspace*{-1em}
\section{Conclusion}

In this paper, we presented \polytest a novel approach for the challenge of LLM-based test generation leveraging the the inherent diversity of multilingual/polyglot capabilities and the creative potential of high-temperature sampling. 
\polytest generates and amplifies a set of tests across multiple languages or through multiple generations, and then unifies these sets, resulting in a significantly consistent and enriched test suite. 
%
Our evaluation on three LLMs, namley \emph{LLama3-70b}, \emph{GPT-4o} and \emph{GPT-3.5} and on the EvalPlus dataset showed
that \polytest is effective in increasing the size of the test suite and its quality. It improved the obtained tests w.r.t. all metrics we considered, namely number of passing tests, statement and branch coverage, and especially the mutation score that is a better indicator for the tests quality. 
\polytest also outperformed Pynguin as a baseline comparison. 
Based on our findings, we recommend the following. First, developers should be aware that generating tests through multiple iterations or across different languages may introduce contradictory test cases. Such inconsistencies can lead to false-positive alerts or create unwarranted confidence in an actually buggy implementation. Second, when working with languages that have weaker support, it is advisable to switch to a more robust language and translate back. This strategy, which is central to \polytest, can mitigate the risk of generating suboptimal test cases.
Third, even for well-supported languages, \polytest consistently outperforms single-language or single-repetition solutions without requiring on-the-fly execution for every test case. 
 
%
For future work, we plan to extend our evaluation with other languages, such as Ruby, Rust, Swift, Go, etc, and to experiment with generating more than five iterations. One limit of our evaluation is the lack of complex programs that we plan to evaluate on in future. However, this will likely require the adaptation of \polytest to take into account the code implementation of a complex program with its dependencies. This can be trickier to handle than one might think at first glance. 
Finally, we plan to replicate our study but in other contexts and tasks, such as generation of code solutions, of code patches/repairs, etc. 
If observed results and gains of \polytest will be observed in other contexts tasks, our methodology could have a deeper impact on the trust or confidence in the LLM results. 

\section*{Reproduction package} Our implementation and evaluation are available in \cite{replicationpackage}.

\begin{acks}
This work is supported by the Inria Défi LLM4Code.
\end{acks}

\bibliographystyle{ACM-Reference-Format}
\bibliography{main}


\begin{thebibliography}{48}


\ifx \showCODEN    \undefined \def \showCODEN     #1{\unskip}     \fi
\ifx \showDOI      \undefined \def \showDOI       #1{#1}\fi
\ifx \showISBNx    \undefined \def \showISBNx     #1{\unskip}     \fi
\ifx \showISBNxiii \undefined \def \showISBNxiii  #1{\unskip}     \fi
\ifx \showISSN     \undefined \def \showISSN      #1{\unskip}     \fi
\ifx \showLCCN     \undefined \def \showLCCN      #1{\unskip}     \fi
\ifx \shownote     \undefined \def \shownote      #1{#1}          \fi
\ifx \showarticletitle \undefined \def \showarticletitle #1{#1}   \fi
\ifx \showURL      \undefined \def \showURL       {\relax}        \fi
\providecommand\bibfield[2]{#2}
\providecommand\bibinfo[2]{#2}
\providecommand\natexlab[1]{#1}
\providecommand\showeprint[2][]{arXiv:#2}

\bibitem[rep({[n.\,d.]})]%
        {replicationpackage}
 \bibinfo{year}{[n.\,d.]}\natexlab{}.
\newblock \bibinfo{title}{Replication Package}.
\newblock
  \bibinfo{howpublished}{\url{https://anonymous.4open.science/r/Polytest-C96C/}}.
\newblock
\newblock
\shownote{Accessed: 2025-03-14}.


\bibitem[Abukhalaf et~al\mbox{.}(2023a)]%
        {10173990}
\bibfield{author}{\bibinfo{person}{Seif Abukhalaf}, \bibinfo{person}{Mohammad
  Hamdaqa}, {and} \bibinfo{person}{Foutse Khomh}.}
  \bibinfo{year}{2023}\natexlab{a}.
\newblock \showarticletitle{On Codex Prompt Engineering for OCL Generation: An
  Empirical Study}. In \bibinfo{booktitle}{\emph{2023 IEEE/ACM 20th
  International Conference on Mining Software Repositories (MSR)}}.
  \bibinfo{pages}{148--157}.
\newblock
\urldef\tempurl%
\url{https://doi.org/10.1109/MSR59073.2023.00033}
\showDOI{\tempurl}


\bibitem[Abukhalaf et~al\mbox{.}(2023b)]%
        {AbukhalafHK23}
\bibfield{author}{\bibinfo{person}{Seif Abukhalaf}, \bibinfo{person}{Mohammad
  Hamdaqa}, {and} \bibinfo{person}{Foutse Khomh}.}
  \bibinfo{year}{2023}\natexlab{b}.
\newblock \showarticletitle{On Codex Prompt Engineering for {OCL} Generation:
  An Empirical Study}. In \bibinfo{booktitle}{\emph{20th {IEEE/ACM}
  International Conference on Mining Software Repositories, {MSR} 2023,
  Melbourne, Australia, May 15-16, 2023}}. \bibinfo{publisher}{{IEEE}},
  \bibinfo{pages}{148--157}.
\newblock
\urldef\tempurl%
\url{https://doi.org/10.1109/MSR59073.2023.00033}
\showDOI{\tempurl}


\bibitem[Andrews et~al\mbox{.}(2006)]%
        {andrews2006using}
\bibfield{author}{\bibinfo{person}{James~H Andrews}, \bibinfo{person}{Lionel~C
  Briand}, \bibinfo{person}{Yvan Labiche}, {and} \bibinfo{person}{Akbar~Siami
  Namin}.} \bibinfo{year}{2006}\natexlab{}.
\newblock \showarticletitle{Using mutation analysis for assessing and comparing
  testing coverage criteria}.
\newblock \bibinfo{journal}{\emph{IEEE Transactions on Software Engineering}}
  \bibinfo{volume}{32}, \bibinfo{number}{8} (\bibinfo{year}{2006}),
  \bibinfo{pages}{608--624}.
\newblock


\bibitem[Arteca et~al\mbox{.}(2022)]%
        {10.1145/3510003.3510106}
\bibfield{author}{\bibinfo{person}{Ellen Arteca}, \bibinfo{person}{Sebastian
  Harner}, \bibinfo{person}{Michael Pradel}, {and} \bibinfo{person}{Frank
  Tip}.} \bibinfo{year}{2022}\natexlab{}.
\newblock \showarticletitle{Nessie: automatically testing JavaScript APIs with
  asynchronous callbacks}. In \bibinfo{booktitle}{\emph{Proceedings of the 44th
  International Conference on Software Engineering}} (Pittsburgh, Pennsylvania)
  \emph{(\bibinfo{series}{ICSE '22})}. \bibinfo{publisher}{Association for
  Computing Machinery}, \bibinfo{address}{New York, NY, USA},
  \bibinfo{pages}{1494–1505}.
\newblock
\showISBNx{9781450392211}
\urldef\tempurl%
\url{https://doi.org/10.1145/3510003.3510106}
\showDOI{\tempurl}


\bibitem[Austin et~al\mbox{.}(2021)]%
        {austin2021program}
\bibfield{author}{\bibinfo{person}{Jacob Austin}, \bibinfo{person}{Augustus
  Odena}, \bibinfo{person}{Maxwell Nye}, \bibinfo{person}{Maarten Bosma},
  \bibinfo{person}{Henryk Michalewski}, \bibinfo{person}{David Dohan},
  \bibinfo{person}{Ellen Jiang}, \bibinfo{person}{Carrie Cai},
  \bibinfo{person}{Michael Terry}, \bibinfo{person}{Quoc Le}, {et~al\mbox{.}}}
  \bibinfo{year}{2021}\natexlab{}.
\newblock \showarticletitle{Program synthesis with large language models}.
\newblock \bibinfo{journal}{\emph{arXiv preprint arXiv:2108.07732}}
  (\bibinfo{year}{2021}).
\newblock


\bibitem[Barei{\ss} et~al\mbox{.}(2022)]%
        {bareiss2022code}
\bibfield{author}{\bibinfo{person}{Patrick Barei{\ss}},
  \bibinfo{person}{Beatriz Souza}, \bibinfo{person}{Marcelo d'Amorim}, {and}
  \bibinfo{person}{Michael Pradel}.} \bibinfo{year}{2022}\natexlab{}.
\newblock \showarticletitle{Code generation tools (almost) for free? a study of
  few-shot, pre-trained language models on code}.
\newblock \bibinfo{journal}{\emph{arXiv preprint arXiv:2206.01335}}
  (\bibinfo{year}{2022}).
\newblock


\bibitem[Baudry et~al\mbox{.}(2024)]%
        {baudry2024generative}
\bibfield{author}{\bibinfo{person}{Benoit Baudry}, \bibinfo{person}{Khashayar
  Etemadi}, \bibinfo{person}{Sen Fang}, \bibinfo{person}{Yogya Gamage},
  \bibinfo{person}{Yi Liu}, \bibinfo{person}{Yuxin Liu},
  \bibinfo{person}{Martin Monperrus}, \bibinfo{person}{Javier Ron},
  \bibinfo{person}{Andr{\'e} Silva}, {and} \bibinfo{person}{Deepika Tiwari}.}
  \bibinfo{year}{2024}\natexlab{}.
\newblock \showarticletitle{Generative AI to Generate Test Data Generators}.
\newblock \bibinfo{journal}{\emph{arXiv preprint arXiv:2401.17626}}
  (\bibinfo{year}{2024}).
\newblock


\bibitem[C{\'a}mara et~al\mbox{.}(2023)]%
        {camara2023assessment}
\bibfield{author}{\bibinfo{person}{Javier C{\'a}mara}, \bibinfo{person}{Javier
  Troya}, \bibinfo{person}{Lola Burgue{\~n}o}, {and} \bibinfo{person}{Antonio
  Vallecillo}.} \bibinfo{year}{2023}\natexlab{}.
\newblock \showarticletitle{On the assessment of generative AI in modeling
  tasks: an experience report with ChatGPT and UML}.
\newblock \bibinfo{journal}{\emph{Software and Systems Modeling}}
  (\bibinfo{year}{2023}), \bibinfo{pages}{1--13}.
\newblock


\bibitem[Chaaben et~al\mbox{.}(2023)]%
        {chaaben2023towards}
\bibfield{author}{\bibinfo{person}{Meriem~Ben Chaaben}, \bibinfo{person}{Lola
  Burgue{\~n}o}, {and} \bibinfo{person}{Houari Sahraoui}.}
  \bibinfo{year}{2023}\natexlab{}.
\newblock \showarticletitle{Towards using few-shot prompt learning for
  automating model completion}. In \bibinfo{booktitle}{\emph{2023 IEEE/ACM 45th
  International Conference on Software Engineering: New Ideas and Emerging
  Results (ICSE-NIER)}}. IEEE, \bibinfo{pages}{7--12}.
\newblock


\bibitem[Chen et~al\mbox{.}(2022)]%
        {chen2022codet}
\bibfield{author}{\bibinfo{person}{Bei Chen}, \bibinfo{person}{Fengji Zhang},
  \bibinfo{person}{Anh Nguyen}, \bibinfo{person}{Daoguang Zan},
  \bibinfo{person}{Zeqi Lin}, \bibinfo{person}{Jian-Guang Lou}, {and}
  \bibinfo{person}{Weizhu Chen}.} \bibinfo{year}{2022}\natexlab{}.
\newblock \showarticletitle{Codet: Code generation with generated tests}.
\newblock \bibinfo{journal}{\emph{arXiv preprint arXiv:2207.10397}}
  (\bibinfo{year}{2022}).
\newblock


\bibitem[Chen et~al\mbox{.}(2023)]%
        {10344012}
\bibfield{author}{\bibinfo{person}{K. Chen}, \bibinfo{person}{Y. Yang},
  \bibinfo{person}{B. Chen}, \bibinfo{person}{J.~Hernandez Lopez},
  \bibinfo{person}{G. Mussbacher}, {and} \bibinfo{person}{D. Varro}.}
  \bibinfo{year}{2023}\natexlab{}.
\newblock \showarticletitle{Automated Domain Modeling with Large Language
  Models: A Comparative Study}. In \bibinfo{booktitle}{\emph{2023 ACM/IEEE 26th
  International Conference on Model Driven Engineering Languages and Systems
  (MODELS)}}. \bibinfo{publisher}{IEEE Computer Society}, \bibinfo{address}{Los
  Alamitos, CA, USA}, \bibinfo{pages}{162--172}.
\newblock
\urldef\tempurl%
\url{https://doi.org/10.1109/MODELS58315.2023.00037}
\showDOI{\tempurl}


\bibitem[Chen et~al\mbox{.}(2021)]%
        {chen2021evaluating}
\bibfield{author}{\bibinfo{person}{Mark Chen}, \bibinfo{person}{Jerry Tworek},
  \bibinfo{person}{Heewoo Jun}, \bibinfo{person}{Qiming Yuan},
  \bibinfo{person}{Henrique Ponde de~Oliveira Pinto}, \bibinfo{person}{Jared
  Kaplan}, \bibinfo{person}{Harri Edwards}, \bibinfo{person}{Yuri Burda},
  \bibinfo{person}{Nicholas Joseph}, \bibinfo{person}{Greg Brockman},
  {et~al\mbox{.}}} \bibinfo{year}{2021}\natexlab{}.
\newblock \showarticletitle{Evaluating large language models trained on code}.
\newblock \bibinfo{journal}{\emph{arXiv preprint arXiv:2107.03374}}
  (\bibinfo{year}{2021}).
\newblock


\bibitem[Dakhel et~al\mbox{.}(2024)]%
        {dakhel2024effective}
\bibfield{author}{\bibinfo{person}{Arghavan~Moradi Dakhel},
  \bibinfo{person}{Amin Nikanjam}, \bibinfo{person}{Vahid Majdinasab},
  \bibinfo{person}{Foutse Khomh}, {and} \bibinfo{person}{Michel~C Desmarais}.}
  \bibinfo{year}{2024}\natexlab{}.
\newblock \showarticletitle{Effective test generation using pre-trained large
  language models and mutation testing}.
\newblock \bibinfo{journal}{\emph{Information and Software Technology}}
  \bibinfo{volume}{171} (\bibinfo{year}{2024}), \bibinfo{pages}{107468}.
\newblock


\bibitem[D{\"o}derlein et~al\mbox{.}(2022)]%
        {doderlein2022piloting}
\bibfield{author}{\bibinfo{person}{Jean-Baptiste D{\"o}derlein},
  \bibinfo{person}{Mathieu Acher}, \bibinfo{person}{Djamel~Eddine Khelladi},
  {and} \bibinfo{person}{Benoit Combemale}.} \bibinfo{year}{2022}\natexlab{}.
\newblock \showarticletitle{Piloting Copilot and Codex: Hot Temperature, Cold
  Prompts, or Black Magic?}
\newblock \bibinfo{journal}{\emph{arXiv preprint arXiv:2210.14699}}
  (\bibinfo{year}{2022}).
\newblock


\bibitem[El~Haji et~al\mbox{.}(2024)]%
        {el2024using}
\bibfield{author}{\bibinfo{person}{Khalid El~Haji}, \bibinfo{person}{Carolin
  Brandt}, {and} \bibinfo{person}{Andy Zaidman}.}
  \bibinfo{year}{2024}\natexlab{}.
\newblock \showarticletitle{Using GitHub Copilot for Test Generation in Python:
  An Empirical Study}.
\newblock  (\bibinfo{year}{2024}).
\newblock


\bibitem[Fraser and Arcuri(2011)]%
        {10.1145/2025113.2025179}
\bibfield{author}{\bibinfo{person}{Gordon Fraser} {and} \bibinfo{person}{Andrea
  Arcuri}.} \bibinfo{year}{2011}\natexlab{}.
\newblock \showarticletitle{EvoSuite: automatic test suite generation for
  object-oriented software}. In \bibinfo{booktitle}{\emph{Proceedings of the
  19th ACM SIGSOFT Symposium and the 13th European Conference on Foundations of
  Software Engineering}} (Szeged, Hungary) \emph{(\bibinfo{series}{ESEC/FSE
  '11})}. \bibinfo{publisher}{Association for Computing Machinery},
  \bibinfo{address}{New York, NY, USA}, \bibinfo{pages}{416–419}.
\newblock
\showISBNx{9781450304436}
\urldef\tempurl%
\url{https://doi.org/10.1145/2025113.2025179}
\showDOI{\tempurl}


\bibitem[Fu et~al\mbox{.}(2023)]%
        {fu2023chatgpt}
\bibfield{author}{\bibinfo{person}{Michael Fu}, \bibinfo{person}{Chakkrit
  Tantithamthavorn}, \bibinfo{person}{Van Nguyen}, {and} \bibinfo{person}{Trung
  Le}.} \bibinfo{year}{2023}\natexlab{}.
\newblock \showarticletitle{ChatGPT for Vulnerability Detection,
  Classification, and Repair: How Far Are We?}
\newblock \bibinfo{journal}{\emph{arXiv preprint arXiv:2310.09810}}
  (\bibinfo{year}{2023}).
\newblock


\bibitem[Gu et~al\mbox{.}(2024)]%
        {gu2024testart}
\bibfield{author}{\bibinfo{person}{Siqi Gu}, \bibinfo{person}{Chunrong Fang},
  \bibinfo{person}{Quanjun Zhang}, \bibinfo{person}{Fangyuan Tian}, {and}
  \bibinfo{person}{Zhenyu Chen}.} \bibinfo{year}{2024}\natexlab{}.
\newblock \showarticletitle{Testart: Improving llm-based unit test via
  co-evolution of automated generation and repair iteration}.
\newblock \bibinfo{journal}{\emph{arXiv e-prints}} (\bibinfo{year}{2024}),
  \bibinfo{pages}{arXiv--2408}.
\newblock


\bibitem[Guo et~al\mbox{.}(2023)]%
        {guo2023exploring}
\bibfield{author}{\bibinfo{person}{Qi Guo}, \bibinfo{person}{Junming Cao},
  \bibinfo{person}{Xiaofei Xie}, \bibinfo{person}{Shangqing Liu},
  \bibinfo{person}{Xiaohong Li}, \bibinfo{person}{Bihuan Chen}, {and}
  \bibinfo{person}{Xin Peng}.} \bibinfo{year}{2023}\natexlab{}.
\newblock \showarticletitle{Exploring the Potential of ChatGPT in Automated
  Code Refinement: An Empirical Study}.
\newblock \bibinfo{journal}{\emph{arXiv preprint arXiv:2309.08221}}
  (\bibinfo{year}{2023}).
\newblock


\bibitem[Hou et~al\mbox{.}(2023)]%
        {hou2023large}
\bibfield{author}{\bibinfo{person}{Xinyi Hou}, \bibinfo{person}{Yanjie Zhao},
  \bibinfo{person}{Yue Liu}, \bibinfo{person}{Zhou Yang},
  \bibinfo{person}{Kailong Wang}, \bibinfo{person}{Li Li},
  \bibinfo{person}{Xiapu Luo}, \bibinfo{person}{David Lo},
  \bibinfo{person}{John Grundy}, {and} \bibinfo{person}{Haoyu Wang}.}
  \bibinfo{year}{2023}\natexlab{}.
\newblock \showarticletitle{Large language models for software engineering: A
  systematic literature review}.
\newblock \bibinfo{journal}{\emph{arXiv preprint arXiv:2308.10620}}
  (\bibinfo{year}{2023}).
\newblock


\bibitem[Jiang et~al\mbox{.}(2023)]%
        {jiang2023selfevolve}
\bibfield{author}{\bibinfo{person}{Shuyang Jiang}, \bibinfo{person}{Yuhao
  Wang}, {and} \bibinfo{person}{Yu Wang}.} \bibinfo{year}{2023}\natexlab{}.
\newblock \showarticletitle{SelfEvolve: A Code Evolution Framework via Large
  Language Models}.
\newblock \bibinfo{journal}{\emph{arXiv preprint arXiv:2306.02907}}
  (\bibinfo{year}{2023}).
\newblock


\bibitem[Kabir et~al\mbox{.}(2023)]%
        {kabir2023empirical}
\bibfield{author}{\bibinfo{person}{Md~Mahir~Asef Kabir},
  \bibinfo{person}{Sk~Adnan Hassan}, \bibinfo{person}{Xiaoyin Wang},
  \bibinfo{person}{Ying Wang}, \bibinfo{person}{Hai Yu}, {and}
  \bibinfo{person}{Na Meng}.} \bibinfo{year}{2023}\natexlab{}.
\newblock \showarticletitle{An empirical study of ChatGPT-3.5 on question
  answering and code maintenance}.
\newblock \bibinfo{journal}{\emph{arXiv preprint arXiv:2310.02104}}
  (\bibinfo{year}{2023}).
\newblock


\bibitem[Lahiri et~al\mbox{.}(2022)]%
        {lahiri2022interactive}
\bibfield{author}{\bibinfo{person}{Shuvendu~K Lahiri}, \bibinfo{person}{Aaditya
  Naik}, \bibinfo{person}{Georgios Sakkas}, \bibinfo{person}{Piali Choudhury},
  \bibinfo{person}{Curtis von Veh}, \bibinfo{person}{Madanlal Musuvathi},
  \bibinfo{person}{Jeevana~Priya Inala}, \bibinfo{person}{Chenglong Wang},
  {and} \bibinfo{person}{Jianfeng Gao}.} \bibinfo{year}{2022}\natexlab{}.
\newblock \showarticletitle{Interactive code generation via test-driven
  user-intent formalization}.
\newblock \bibinfo{journal}{\emph{arXiv preprint arXiv:2208.05950}}
  (\bibinfo{year}{2022}).
\newblock


\bibitem[Lemieux et~al\mbox{.}(2023)]%
        {lemieux2023codamosa}
\bibfield{author}{\bibinfo{person}{Caroline Lemieux},
  \bibinfo{person}{Jeevana~Priya Inala}, \bibinfo{person}{Shuvendu~K Lahiri},
  {and} \bibinfo{person}{Siddhartha Sen}.} \bibinfo{year}{2023}\natexlab{}.
\newblock \showarticletitle{Codamosa: Escaping coverage plateaus in test
  generation with pre-trained large language models}. In
  \bibinfo{booktitle}{\emph{2023 IEEE/ACM 45th International Conference on
  Software Engineering (ICSE)}}. IEEE, \bibinfo{pages}{919--931}.
\newblock


\bibitem[Li and Yuan(2024)]%
        {li2024large}
\bibfield{author}{\bibinfo{person}{Kefan Li} {and} \bibinfo{person}{Yuan
  Yuan}.} \bibinfo{year}{2024}\natexlab{}.
\newblock \showarticletitle{Large Language Models as Test Case Generators:
  Performance Evaluation and Enhancement}.
\newblock \bibinfo{journal}{\emph{arXiv preprint arXiv:2404.13340}}
  (\bibinfo{year}{2024}).
\newblock


\bibitem[Li et~al\mbox{.}(2009)]%
        {li2009experimental}
\bibfield{author}{\bibinfo{person}{Nan Li}, \bibinfo{person}{Upsorn
  Praphamontripong}, {and} \bibinfo{person}{Jeff Offutt}.}
  \bibinfo{year}{2009}\natexlab{}.
\newblock \showarticletitle{An experimental comparison of four unit test
  criteria: Mutation, edge-pair, all-uses and prime path coverage}. In
  \bibinfo{booktitle}{\emph{2009 International Conference on Software Testing,
  Verification, and Validation Workshops}}. IEEE, \bibinfo{pages}{220--229}.
\newblock


\bibitem[Liu et~al\mbox{.}(2023)]%
        {liu2023improving}
\bibfield{author}{\bibinfo{person}{Chao Liu}, \bibinfo{person}{Xuanlin Bao},
  \bibinfo{person}{Hongyu Zhang}, \bibinfo{person}{Neng Zhang},
  \bibinfo{person}{Haibo Hu}, \bibinfo{person}{Xiaohong Zhang}, {and}
  \bibinfo{person}{Meng Yan}.} \bibinfo{year}{2023}\natexlab{}.
\newblock \showarticletitle{Improving ChatGPT Prompt for Code Generation}.
\newblock \bibinfo{journal}{\emph{arXiv preprint arXiv:2305.08360}}
  (\bibinfo{year}{2023}).
\newblock


\bibitem[Liu et~al\mbox{.}(2024)]%
        {liu2024your}
\bibfield{author}{\bibinfo{person}{Jiawei Liu}, \bibinfo{person}{Chunqiu~Steven
  Xia}, \bibinfo{person}{Yuyao Wang}, {and} \bibinfo{person}{Lingming Zhang}.}
  \bibinfo{year}{2024}\natexlab{}.
\newblock \showarticletitle{Is your code generated by chatgpt really correct?
  rigorous evaluation of large language models for code generation}.
\newblock \bibinfo{journal}{\emph{Advances in Neural Information Processing
  Systems}}  \bibinfo{volume}{36} (\bibinfo{year}{2024}).
\newblock


\bibitem[Lukasczyk et~al\mbox{.}(2023)]%
        {lukasczyk2023empirical}
\bibfield{author}{\bibinfo{person}{Stephan Lukasczyk}, \bibinfo{person}{Florian
  Kroi{\ss}}, {and} \bibinfo{person}{Gordon Fraser}.}
  \bibinfo{year}{2023}\natexlab{}.
\newblock \showarticletitle{An empirical study of automated unit test
  generation for Python}.
\newblock \bibinfo{journal}{\emph{Empirical Software Engineering}}
  \bibinfo{volume}{28}, \bibinfo{number}{2} (\bibinfo{year}{2023}),
  \bibinfo{pages}{36}.
\newblock


\bibitem[Nashid et~al\mbox{.}(2023)]%
        {nashid2023retrieval}
\bibfield{author}{\bibinfo{person}{Noor Nashid}, \bibinfo{person}{Mifta
  Sintaha}, {and} \bibinfo{person}{Ali Mesbah}.}
  \bibinfo{year}{2023}\natexlab{}.
\newblock \showarticletitle{Retrieval-based prompt selection for code-related
  few-shot learning}. In \bibinfo{booktitle}{\emph{2023 IEEE/ACM 45th
  International Conference on Software Engineering (ICSE)}}. IEEE,
  \bibinfo{pages}{2450--2462}.
\newblock


\bibitem[Nathalia et~al\mbox{.}(2023)]%
        {nathalia2023artificial}
\bibfield{author}{\bibinfo{person}{Nascimento Nathalia},
  \bibinfo{person}{Alencar Paulo}, {and} \bibinfo{person}{Cowan Donald}.}
  \bibinfo{year}{2023}\natexlab{}.
\newblock \showarticletitle{Artificial Intelligence vs. Software Engineers: An
  Empirical Study on Performance and Efficiency using ChatGPT}. In
  \bibinfo{booktitle}{\emph{Proceedings of the 33rd Annual International
  Conference on Computer Science and Software Engineering}}.
  \bibinfo{pages}{24--33}.
\newblock


\bibitem[Nguyen and Nadi(2022)]%
        {nguyen2022empirical}
\bibfield{author}{\bibinfo{person}{Nhan Nguyen} {and} \bibinfo{person}{Sarah
  Nadi}.} \bibinfo{year}{2022}\natexlab{}.
\newblock \showarticletitle{An empirical evaluation of GitHub copilot's code
  suggestions}. In \bibinfo{booktitle}{\emph{Proceedings of the 19th
  International Conference on Mining Software Repositories}}.
  \bibinfo{pages}{1--5}.
\newblock


\bibitem[Ozkaya(2023)]%
        {10109345}
\bibfield{author}{\bibinfo{person}{Ipek Ozkaya}.}
  \bibinfo{year}{2023}\natexlab{}.
\newblock \showarticletitle{Application of Large Language Models to Software
  Engineering Tasks: Opportunities, Risks, and Implications}.
\newblock \bibinfo{journal}{\emph{IEEE Software}} \bibinfo{volume}{40},
  \bibinfo{number}{3} (\bibinfo{year}{2023}), \bibinfo{pages}{4--8}.
\newblock
\urldef\tempurl%
\url{https://doi.org/10.1109/MS.2023.3248401}
\showDOI{\tempurl}


\bibitem[Pan et~al\mbox{.}(2024)]%
        {pan2024multi}
\bibfield{author}{\bibinfo{person}{Rangeet Pan}, \bibinfo{person}{Myeongsoo
  Kim}, \bibinfo{person}{Rahul Krishna}, \bibinfo{person}{Raju Pavuluri}, {and}
  \bibinfo{person}{Saurabh Sinha}.} \bibinfo{year}{2024}\natexlab{}.
\newblock \showarticletitle{ASTER: Natural and Multi-language Unit Test
  Generation with LLMs}.
\newblock \bibinfo{journal}{\emph{arXiv preprint arXiv:2409.03093}}
  (\bibinfo{year}{2024}).
\newblock


\bibitem[Parsai and Demeyer(2020)]%
        {parsai2020comparing}
\bibfield{author}{\bibinfo{person}{Ali Parsai} {and} \bibinfo{person}{Serge
  Demeyer}.} \bibinfo{year}{2020}\natexlab{}.
\newblock \showarticletitle{Comparing mutation coverage against branch coverage
  in an industrial setting}.
\newblock \bibinfo{journal}{\emph{International Journal on Software Tools for
  Technology Transfer}} \bibinfo{volume}{22}, \bibinfo{number}{4}
  (\bibinfo{year}{2020}), \bibinfo{pages}{365--388}.
\newblock


\bibitem[Pearce et~al\mbox{.}(2022)]%
        {pearce2022asleep}
\bibfield{author}{\bibinfo{person}{Hammond Pearce}, \bibinfo{person}{Baleegh
  Ahmad}, \bibinfo{person}{Benjamin Tan}, \bibinfo{person}{Brendan
  Dolan-Gavitt}, {and} \bibinfo{person}{Ramesh Karri}.}
  \bibinfo{year}{2022}\natexlab{}.
\newblock \showarticletitle{Asleep at the keyboard? assessing the security of
  github copilot’s code contributions}. In \bibinfo{booktitle}{\emph{2022
  IEEE Symposium on Security and Privacy (SP)}}. IEEE,
  \bibinfo{pages}{754--768}.
\newblock


\bibitem[Sapozhnikov et~al\mbox{.}(2024)]%
        {sapozhnikov2024testspark}
\bibfield{author}{\bibinfo{person}{Arkadii Sapozhnikov},
  \bibinfo{person}{Mitchell Olsthoorn}, \bibinfo{person}{Annibale Panichella},
  \bibinfo{person}{Vladimir Kovalenko}, {and} \bibinfo{person}{Pouria
  Derakhshanfar}.} \bibinfo{year}{2024}\natexlab{}.
\newblock \showarticletitle{TestSpark: IntelliJ IDEA's Ultimate Test Generation
  Companion}.
\newblock \bibinfo{journal}{\emph{arXiv preprint arXiv:2401.06580}}
  (\bibinfo{year}{2024}).
\newblock


\bibitem[Sch{\"a}fer et~al\mbox{.}(2023)]%
        {schafer2023empirical}
\bibfield{author}{\bibinfo{person}{Max Sch{\"a}fer}, \bibinfo{person}{Sarah
  Nadi}, \bibinfo{person}{Aryaz Eghbali}, {and} \bibinfo{person}{Frank Tip}.}
  \bibinfo{year}{2023}\natexlab{}.
\newblock \showarticletitle{An empirical evaluation of using large language
  models for automated unit test generation}.
\newblock \bibinfo{journal}{\emph{IEEE Transactions on Software Engineering}}
  (\bibinfo{year}{2023}).
\newblock


\bibitem[Siddiq et~al\mbox{.}(2024)]%
        {siddiq2024using}
\bibfield{author}{\bibinfo{person}{Mohammed~Latif Siddiq},
  \bibinfo{person}{Joanna~CS Santos}, \bibinfo{person}{Ridwanul~Hasan Tanvir},
  \bibinfo{person}{Noshin Ulfat}, \bibinfo{person}{Fahmid Al~Rifat}, {and}
  \bibinfo{person}{Vin{\'\i}cius~Carvalho Lopes}.}
  \bibinfo{year}{2024}\natexlab{}.
\newblock \showarticletitle{Using Large Language Models to Generate JUnit
  Tests: An Empirical Study}.
\newblock  (\bibinfo{year}{2024}).
\newblock


\bibitem[Sobania et~al\mbox{.}(2022)]%
        {sobania2022choose}
\bibfield{author}{\bibinfo{person}{Dominik Sobania}, \bibinfo{person}{Martin
  Briesch}, {and} \bibinfo{person}{Franz Rothlauf}.}
  \bibinfo{year}{2022}\natexlab{}.
\newblock \showarticletitle{Choose your programming copilot: A comparison of
  the program synthesis performance of github copilot and genetic programming}.
  In \bibinfo{booktitle}{\emph{Proceedings of the genetic and evolutionary
  computation conference}}. \bibinfo{pages}{1019--1027}.
\newblock


\bibitem[Vaithilingam et~al\mbox{.}(2022)]%
        {vaithilingam2022expectation}
\bibfield{author}{\bibinfo{person}{Priyan Vaithilingam},
  \bibinfo{person}{Tianyi Zhang}, {and} \bibinfo{person}{Elena~L Glassman}.}
  \bibinfo{year}{2022}\natexlab{}.
\newblock \showarticletitle{Expectation vs. experience: Evaluating the
  usability of code generation tools powered by large language models}. In
  \bibinfo{booktitle}{\emph{Chi conference on human factors in computing
  systems extended abstracts}}. \bibinfo{pages}{1--7}.
\newblock


\bibitem[Wang et~al\mbox{.}(2024)]%
        {wang2024software}
\bibfield{author}{\bibinfo{person}{Junjie Wang}, \bibinfo{person}{Yuchao
  Huang}, \bibinfo{person}{Chunyang Chen}, \bibinfo{person}{Zhe Liu},
  \bibinfo{person}{Song Wang}, {and} \bibinfo{person}{Qing Wang}.}
  \bibinfo{year}{2024}\natexlab{}.
\newblock \showarticletitle{Software testing with large language models:
  Survey, landscape, and vision}.
\newblock \bibinfo{journal}{\emph{IEEE Transactions on Software Engineering}}
  (\bibinfo{year}{2024}).
\newblock


\bibitem[Wang et~al\mbox{.}(2022)]%
        {wang2022self}
\bibfield{author}{\bibinfo{person}{Xuezhi Wang}, \bibinfo{person}{Jason Wei},
  \bibinfo{person}{Dale Schuurmans}, \bibinfo{person}{Quoc Le},
  \bibinfo{person}{Ed Chi}, \bibinfo{person}{Sharan Narang},
  \bibinfo{person}{Aakanksha Chowdhery}, {and} \bibinfo{person}{Denny Zhou}.}
  \bibinfo{year}{2022}\natexlab{}.
\newblock \showarticletitle{Self-consistency improves chain of thought
  reasoning in language models}.
\newblock \bibinfo{journal}{\emph{arXiv preprint arXiv:2203.11171}}
  (\bibinfo{year}{2022}).
\newblock


\bibitem[Wohlin et~al\mbox{.}(2012)]%
        {wohlin2012experimentation}
\bibfield{author}{\bibinfo{person}{Claes Wohlin}, \bibinfo{person}{Per
  Runeson}, \bibinfo{person}{Martin H{\"o}st}, \bibinfo{person}{Magnus~C
  Ohlsson}, \bibinfo{person}{Bj{\"o}rn Regnell}, {and} \bibinfo{person}{Anders
  Wessl{\'e}n}.} \bibinfo{year}{2012}\natexlab{}.
\newblock \bibinfo{booktitle}{\emph{Experimentation in software engineering}}.
\newblock \bibinfo{publisher}{Springer Science \& Business Media}.
\newblock


\bibitem[Yeti{\c{s}}tiren et~al\mbox{.}(2023)]%
        {yeticstiren2023evaluating}
\bibfield{author}{\bibinfo{person}{Burak Yeti{\c{s}}tiren},
  \bibinfo{person}{I{\c{s}}{\i}k {\"O}zsoy}, \bibinfo{person}{Miray Ayerdem},
  {and} \bibinfo{person}{Eray T{\"u}z{\"u}n}.} \bibinfo{year}{2023}\natexlab{}.
\newblock \showarticletitle{Evaluating the Code Quality of AI-Assisted Code
  Generation Tools: An Empirical Study on GitHub Copilot, Amazon CodeWhisperer,
  and ChatGPT}.
\newblock \bibinfo{journal}{\emph{arXiv preprint arXiv:2304.10778}}
  (\bibinfo{year}{2023}).
\newblock


\bibitem[Zhang et~al\mbox{.}(2023)]%
        {zhang2023multilingual}
\bibfield{author}{\bibinfo{person}{Jiyang Zhang}, \bibinfo{person}{Pengyu Nie},
  \bibinfo{person}{Junyi~Jessy Li}, {and} \bibinfo{person}{Milos Gligoric}.}
  \bibinfo{year}{2023}\natexlab{}.
\newblock \showarticletitle{Multilingual code co-evolution using large language
  models}.
\newblock \bibinfo{journal}{\emph{arXiv preprint arXiv:2307.14991}}
  (\bibinfo{year}{2023}).
\newblock


\bibitem[Ziegler et~al\mbox{.}(2022)]%
        {ziegler2022productivity}
\bibfield{author}{\bibinfo{person}{Albert Ziegler}, \bibinfo{person}{Eirini
  Kalliamvakou}, \bibinfo{person}{X~Alice Li}, \bibinfo{person}{Andrew Rice},
  \bibinfo{person}{Devon Rifkin}, \bibinfo{person}{Shawn Simister},
  \bibinfo{person}{Ganesh Sittampalam}, {and} \bibinfo{person}{Edward
  Aftandilian}.} \bibinfo{year}{2022}\natexlab{}.
\newblock \showarticletitle{Productivity assessment of neural code completion}.
  In \bibinfo{booktitle}{\emph{Proceedings of the 6th ACM SIGPLAN International
  Symposium on Machine Programming}}. \bibinfo{pages}{21--29}.
\newblock


\end{thebibliography}

\end{document}